\documentclass{article}

\usepackage[preprint]{neurips_2025}

\usepackage[utf8]{inputenc} % allow utf-8 input
\usepackage[T1]{fontenc}    % use 8-bit T1 fonts
\usepackage{hyperref}       % hyperlinks
\usepackage{url}            % simple URL typesetting
\usepackage{booktabs}       % professional-quality tables
\usepackage{amsfonts}       % blackboard math symbols
\usepackage{nicefrac}       % compact symbols for 1/2, etc.
\usepackage{microtype}      % microtypography
\usepackage{xcolor}         % colors
\usepackage{graphicx}
\usepackage{amsmath}
\usepackage{subcaption} % Add this to your preamble if not already
\usepackage{tcolorbox}

\title{Dataset and Benchmark for Enhancing Critical Retained Foreign Object Detection}

\author{
  Yuli Wang$^{1}$, Victoria R. Shi$^{2}$, Liwei Zhou$^{3}$, Richard Chin$^{3}$, Yuwei Dai$^{2}$, Yuanyun Hu$^{3}$, \\ \textbf{Cheng-Yi Li}$^{1}$, \textbf{Haoyue Guan}$^{3}$, \textbf{Jiashu Cheng}$^{2}$,  \textbf{Yu Sun}$^{3}$, \textbf{Cheng Ting Lin}$^{2}$, \\ \textbf{Ihab Kamel}$^{4}$, \textbf{Premal Trivedi}$^{4}$, \textbf{Pamela Johnson}$^{2}$, \textbf{John Eng}$^{2}$, \\ \textbf{Harrison Bai}$^{2}$\\
  $^{1}$Department of Biomedical Engineering, Johns Hopkins University \\
  $^{2}$Department of Radiology and Radiological Science, Johns Hopkins University \\
  $^{3}$Department of Electrical and Computer Science, Johns Hopkins University\\
  $^{4}$Department of Radiology, University of Colorado \\
  \texttt{\{ywang687,vshi1,lzhou75,rchin7,ydai55,hguan16,yhu160\}@jhu.edu} \\
  \texttt{\{ysun214,clin97,pjohnso5,jeng,hbai7\}@jhu.edu} \\
  \texttt{\{ihab.kamel,premal.trivedi\}@cuanschutz.edu}\\
}

\begin{document}

\maketitle

\begin{abstract}
Critical retained foreign objects (RFOs), including surgical instruments like sponges and needles, pose serious patient safety risks and carry significant financial and legal implications for healthcare institutions. Detecting critical RFOs using artificial intelligence remains challenging due to their rarity and the limited availability of chest X-ray datasets that specifically feature critical RFOs cases. Existing datasets only contain non-critical RFOs, like necklace or zipper, further limiting their utility for developing clinically impactful detection algorithms. To address these limitations, we introduce "Hopkins RFOs Bench", the first and largest dataset of its kind, containing 144 chest X-ray images of critical RFO cases collected over 18 years from the Johns Hopkins Health System. Using this dataset, we benchmark several state-of-the-art object detection models, highlighting the need for enhanced detection methodologies for critical RFO cases. Recognizing data scarcity challenges, we further explore image synthetic methods to bridge this gap. We evaluate two advanced synthetic image methods—DeepDRR-RFO, a physics-based method, and RoentGen-RFO, a diffusion-based method—for creating realistic radiographs featuring critical RFOs. Our comprehensive analysis identifies the strengths and limitations of each synthetic method, providing insights into effectively utilizing synthetic data to enhance model training. The Hopkins RFOs Bench and our findings significantly advance the development of reliable, generalizable AI-driven solutions for detecting critical RFOs in clinical chest X-rays. Dataset and benchmark code are open at Github \url{https://github.com/YuliWanghust/RFO_Bench}
\end{abstract}

\section{Introduction}
Retained foreign objects (RFOs), such as sponges, needles, sutures, and other instruments, can have severe consequences for patients, leading to significant financial and legal repercussions for the involved medical institutions. These reportable "never events" are rare, with studies estimating an incidence of approximately 1 to 2 in 10,000 surgeries, with rates as high as 1 in 1000 to 1500 for abdominal surgeries \cite{lincourt2007retained,zejnullahu2017retained}. Although RFOs are typically discussed within the context of a surgical procedure, they can also occur after non-surgical interventions. For example, the incidence of retained guidewires following central line placement is estimated at 0.03\% \cite{mariyaselvam2022central,vannucci2013retained}. The actual number of RFO cases is likely underreported due to low reporting rates, and many patients may be asymptomatic, remaining unaware of the event. 

Recently, AI has emerged as a promising approach to improving RFOs detection in radiographs, with potential benefits in clinical workflow optimization \cite{kawakubo2023deep,asiyanbola2012modified,kufel2023chest,santosh2021generic,wang2024car}. Despite encouraging results from deep neural networks, several critical limitations remain\cite{toruner2024artificial,wang2024enhancing}. Foremost is the scarcity of publicly available datasets; most intraoperative radiographs obtained for procedural guidance are not routinely archived, limiting the development of large-scale datasets. Moreover, existing public datasets (shown in Table \ref{tab:comparison_dataset}) only include non-critical RFOs, such as external instruments, tubes, and zippers, which pose no patient risk. Conversely, critical RFOs like needles and surgical sponges, associated with severe complications, are rarely captured due to their infrequent occurrence, limiting the medical relevance of current datasets. These challenges highlight the urgent need for a large dataset with critical RFOs, new solutions to address data constraints, reliably detect both critical and non-critical RFOs, and ensure generalizability across diverse clinical settings.

Moreover, to overcome the limitation posed by insufficient data, various synthetic radiograph generation methods have been proposed, including physics-based methods such as virtual radiograph generators, and denoising diffusion probabilistic model (DDPM). Physics-based methods simulate realistic radiographs by employing physical modeling of anatomical structures and X-ray interactions, allowing for controlled experimentation with various object placements and imaging conditions. On the other hand, DDPM-based methods generate synthetic images by progressively denoising randomly sampled images, enabling the creation of diverse and realistic data without explicit physical modeling. However, both approaches are significantly constrained by the extreme rarity of critical RFOs cases, and thus far, none have incorporated critical RFOs into their synthetic scope, limiting their practical application for training robust AI algorithms for critical RFOs detection.

Addressing these significant challenges in AI-driven RFOs detection on radiographs, we present the following contributions in this paper:
\begin{itemize}
    \item Development of "Hopkins RFOs Bench," \textbf{the first and largest dataset of its kind}, comprising 144 chest X-ray images containing critical RFOs collected from the Johns Hopkins Health System over the past 18 years.
    \vspace{-3pt}
    \item Benchmarking existing object detection models on our proposed dataset to underscore the necessity for advanced detection methods.
    \vspace{-3pt}
    \item Evaluating two customized synthetic image generation models, DeepDRR-RFO and RoentGen-RFO, for creating images with critical RFOs. We train object detection models using these synthetic images, analyze the strengths and weaknesses of each approach, and provide insights to guide future improvements utilizing our openly accessible dataset.
    \vspace{-3pt}
\end{itemize}

\section{Related Works}

\subsection{Clinical Safety Mechanism for RFOs incidence}
Given the medical complications and medicolegal implications of RFOs, several safety mechanisms have been established to reduce their incidence \cite{webb1988management}. In surgical procedures, operating room staff routinely perform final counts to ensure that no foreign objects are left inside the patient. If a discrepancy in the final counts arises, intraoperative radiographs are obtained to detect any RFOs. However, even with these safety measures, more than 80\% of surgeries involving RFOs have reported correct counts at the conclusion of the procedure \cite{gibbs2007preventable}. When clinical teams fail to detect RFOs during the procedure, they are often discovered incidentally during follow-up imaging studies \cite{takahashi2023characteristics}. Radiologists, however, may miss these objects because they do not usually expect to find them during routine imaging interpretation. 

Regarding imaging techniques for RFOs detection, plain radiographs remain the gold standard. On X-ray, RFOs often appear as radiopacities with associated mass effects, mottled air, or density changes in surrounding soft tissues \cite{kumar2017imaging}. Most RFOs contain radiopaque markers, which enhance their visibility on X-ray. However, these markers can become disfigured within the body, reducing their reliability \cite{o2003imaging}. RFOs without radiopaque markers are often identified through cross-sectional imaging or by detecting radiolucency caused by air trapping \cite{yun2021retained}. It is important to note that false-negative radiographs do occur, with some studies suggesting that intraoperative radiographs may miss up to one-third of RFOs \cite{cima2008incidence}. Additionally, radiologists interpreting these radiographs often face time pressures, particularly after surgery, which can increase the likelihood of errors.

\subsection{Deep Learning Based RFOs Detection and Its Dataset}
Kawakubo et al. \cite{kawakubo2023deep} developed a deep learning model that effectively detected retained surgical items through the post-processing of fused images containing surgical sponges and unremarkable postoperative X-rays. In a proof-of-concept study, Asiyanbola et al. \cite{asiyanbola2012modified} employed a map-seeking circuit for the detection of needles within abdominal X-rays. Similarly, Kufel et al. \cite{kufel2023chest} designed a convolutional neural network (CNN) to identify objects in chest X-ray images using the publicly available National Institutes of Health Chest X-ray (CXR) dataset \cite{wang2017chestx}. Additionally, Santosh et al. \cite{santosh2021generic,zhou2020automatically} applied the YOLOv4 model for non-critical RFOs detection on the same public CXR dataset. 

Although CNNs have demonstrated promising capabilities for detecting and identifying RFOs, several limitations persist. A significant barrier to AI-driven RFO detection is the scarcity of publicly available datasets. Most intraoperative radiographs taken for procedural guidance are not archived, and existing datasets only include non-critical RFOs that pose no risk. Critical RFOs, such as needles and sponges, which can cause severe complications if retained internally, are rarely captured due to their extreme scarcity. Consequently, current public datasets have limited clinical relevance.
These challenges highlight the need for a comprehensive critical RFO dataset and new solutions that can effectively address data scarcity, and accurately detect and localize all types of RFOs.

Among prior contributions, Object-CXR \cite{objectcxr} stands out as a large-scale dataset including over 9,000 chest X-ray images. However, Object-CXR lacks the inclusion of critical RFOs cases, representing a significant gap in realistically mimicking clinical RFOs detection scenarios and limiting the effectiveness of models trained on this dataset for RFOs object detection. Additionally, datasets such as MP-RFOs \cite{hogeweg2013foreign}, CXR-RFOs \cite{kufel2023chest}, and CXR-400 \cite{santosh2021generic} not only lack critical RFOs cases but also are not publicly accessible, rendering them unusable for research purposes. In contrast, our proposed dataset, Hopkins RFOs Bench, directly addresses these limitations by including critical RFOs cases, as well as images with no RFOs and non-interest RFOs, thereby enabling more comprehensive model pre-training and fine-tuning. Furthermore, Hopkins RFOs Bench is approved for public open access by the Johns Hopkins IRB, enhancing its availability for the research community.

\vspace{-5pt}
\begin{table}[h]
  \caption{Hopkins RFOs Bench vs. existing RFOs radiographs image datasets}
  \centering
  \small
  \setlength{\tabcolsep}{4pt}
  \renewcommand{\arraystretch}{0.8}
  \begin{tabular}{cccccccc}
    \toprule
    \multicolumn{1}{c}{\textbf{Dataset}} & \multicolumn{1}{c}{\textbf{Access}} & \multicolumn{3}{c}{\textbf{Categories}} & \multicolumn{1}{c}{\textbf{Counts}} & \multicolumn{2}{c}{\textbf{Task Labels}} \\
    \cmidrule(r){2-2} \cmidrule(r){3-5} \cmidrule(r){6-6} \cmidrule(r){7-8}
    & Open? & No & No-critical & Critical & No. & Image-level & Subject-level \\
    \midrule
    Object-CXR \cite{objectcxr} & Yes & $\checkmark$ & $\checkmark$ & $\times$ & 9,000 & $\checkmark$ & $\checkmark$ \\
    MP-RFOs \cite{hogeweg2013foreign} & No & $\times$ & $\checkmark$ & $\times$ & 256 & $\times$ & $\checkmark$ \\
    Flexsim \cite{andriiashenct} & Yes & $\times$ & $\checkmark$ & $\times$ & 4,032 & $\checkmark$ & $\times$ \\
    CXR-RFOs \cite{kufel2023chest} & No & $\times$ & $\checkmark$ & $\times$ & 30,805 & $\checkmark$ & $\checkmark$ \\
    CXR-400 \cite{santosh2021generic} & No & $\times$ & $\checkmark$ & $\times$ & 400 & $\times$ & $\times$ \\
    Ours & Yes & $\checkmark$ & $\checkmark$ & $\checkmark$ & 144 & $\checkmark$ & $\checkmark$ \\
    \bottomrule
  \end{tabular}
  \label{tab:comparison_dataset}
  \vspace{-10pt}
\end{table}

\subsection{Synthetic Chest X-ray Images Generation}

Recent methods for generating synthetic radiographs primarily fall into two categories: physics-based simulation and data-driven generative modeling. Physics-based methods, such as DeepDRR \cite{unberath2018deepdrr}, segment CT volumes into air, soft tissue, and bone, render 3D models of RFOs, and simulate X-ray physics to produce realistic annotated radiographs. These approaches offer anatomically accurate images with precise control over imaging parameters \cite{unberath2019enabling}. Conversely, data-driven methods utilize DDPMs, which iteratively denoise random inputs to create diverse and high-quality images \cite{huijben2024denoising,khosravi2024synthetically,hosseini2025synthetic}. DDPMs model the data distribution by iteratively denoising random noise, allowing for the creation of diverse and high-quality images \cite{khosravi2024synthetically}. For instance, researchers have applied DDPMs to generate radiographs conditioned on demographic and pathological characteristics, demonstrating improved performance and generalizability of classifiers when synthetic data supplements real datasets \cite{irvin2019chexpert,khosravi2024synthetically}. Additionally, DDPMs have been used to generate synthetic images of specific conditions, such as pneumonia, enhancing the training of diagnostic models in scenarios with limited positive cases \cite{mahaulpatha2024ddpm}. Both strategies provide possibilities to enhance the generalizability and accuracy of AI models in medical imaging.

\section{Cohort Definition and Dataset Composition}

\begin{figure}[h]
  \centering
    \includegraphics[width=0.9\textwidth]{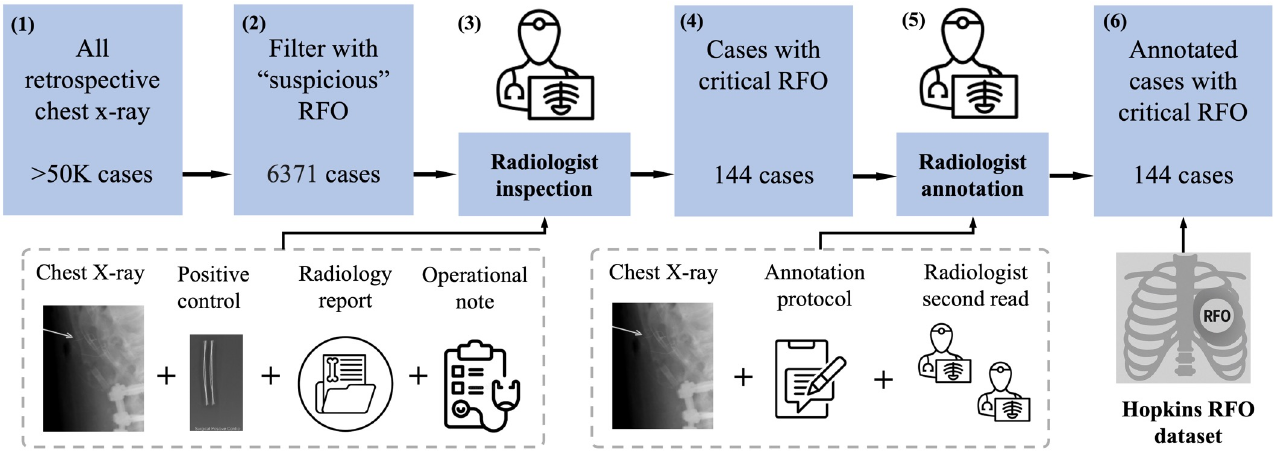}
  \caption{A flowchart outlining our cohort definition process is shown. The steps are as follows: (1) inclusion of all retrospective radiographs from the Hopkins Health System between 2007 and 2024; (2) identification of “suspicious” RFOs cases based on radiology reports containing the terms ‘foreign body’ or ‘radio-opaque foreign body,’ excluding phrases such as ‘no foreign body’ or ‘no radio-opaque foreign body’; (3–4) radiologist inspection of “suspicious” radiographs, using positive RFOs controls, associated radiology reports, and operative notes; (5) radiologist annotation of all types of RFOs—including non-critical and critical cases—using a standardized protocol (details shown in Appendix \ref{anno_prot}) and conducting radiologist secondary review; and (6) assembly of the final dataset: the Hopkins RFOs Bench.}
  \label{fig:cohort}
\end{figure}

Our study, approved by the Johns Hopkins Institutional Review Board (IRB00383214); see Appendix \ref{IRB}), retrospectively identified cases from the Johns Hopkins Health System spanning 2007 to 2024. Utilizing the Johns Hopkins mPower search tool and the Core for Clinical Research Data Acquisition (CCDA) \cite{johnshopkins_ccda}, we initially identified 144 critical RFOs cases. Following our cohort definition protocol, which involved systematic data retrieval, cleaning, and rigorous adherence guided by the radiologists to predefined inclusion criteria, the final cohort consisted of 144 critical RFO cases from 144 distinct patients (see Figure \ref{fig:cohort} for details).

For each patient case, we acquired the original X-ray images in Digital Imaging and Communications in Medicine (DICOM) format, corresponding radiology reports, and operational notes. All data were processed for subsequent analysis and thorough deidentification. Each X-ray image was annotated by trained radiologists according to a standardized protocol, involving both image- and object-level annotations, including polygonal and bounding box delineations (detailed annotation protocol provided in Appendix \ref{anno_prot}). Furthermore, we established training, validation, and testing dataset splits, allocating 70\%, 10\%, and 20\% of the cases, respectively. These splits were strictly patient-based, ensuring that data from each patient appeared exclusively within a single subset.

\begin{figure}[h]
  \centering
    \includegraphics[width=0.8\textwidth]{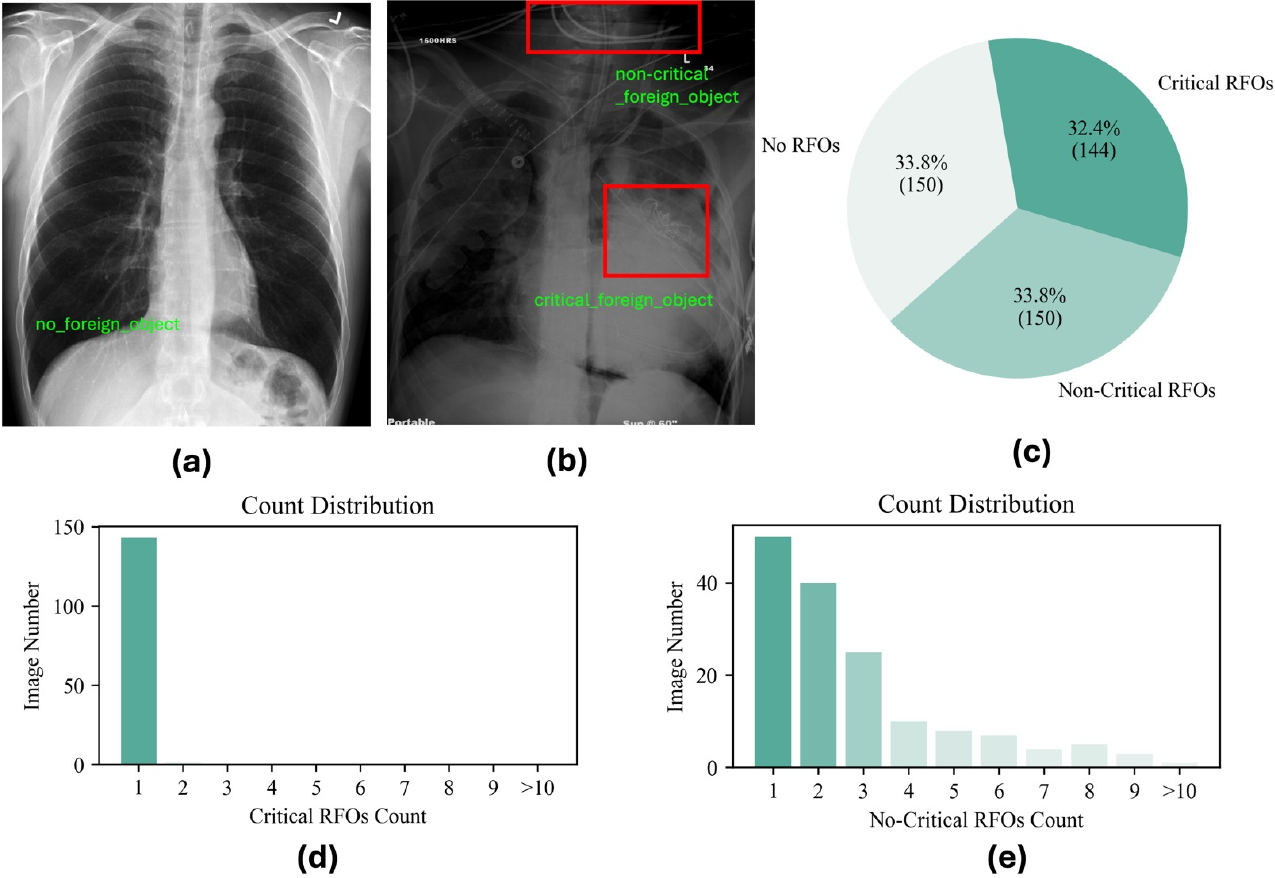}
  \caption{Examples and statistics from the Hopkins RFOs Bench. (a) An annotated chest X-ray image without RFOs (No RFOs image). (b) An annotated chest X-ray image with non-critical and critical RFOs. (c) Dataset composition summary, categorizing 144 critical RFOs cases, 150 No RFOs cases, and 150 No-critical RFOs cases. Distribution of critical RFOs counts per image for critical images (d) and no-critical images (e).}
  \label{fig:statistics}
  \vspace{-5pt}
\end{figure}

Examples of annotated chest X-ray images from our dataset are presented in Figure \ref{fig:statistics} (a) No RFOs image, and (b) critical RFOs image. At the image level, images are categorized into RFOs and non-RFOs groups. At the object level, annotations were performed using bounding boxes or polygonal delineations to identify RFOs objects and classify them into non-critical or critical RFOs categories. Figure \ref{fig:statistics} (c) summarizes the composition of our proposed dataset, comprising 144 critical RFOs cases, 150 No RFOs cases, and 150 No-critical RFOs cases (No RFOs and No-critical RFOs cases are used to construct a class-balanced dataset). Additionally, Figure \ref{fig:statistics} (d) presents the RFOs count distribution, indicating an average of 1.0 critical RFOs per critical RFOs image and 2.7 RFOs per critical RFOs image. Based on this rigorously defined cohort, we publicly release the resulting dataset as the Hopkins RFOs Bench, which includes:

\begin{itemize}
\item Chest X-Ray Images: Imaging slices of chest X-rays in our cohort, provided in JPG format.
%\item DICOM Headers: A subset of headers from the original DICOM files, including anonymized patient ID, anonymized study date, instance order within the series, patient position, pixel spacing and imaging machine manufacturer.
\item Image-Level Label: Identifies whether each chest X-ray image is without or with RFOs (including non-critical or critical RFOs).
\item Object-Level Label: Polygonal or bounding box annotations indicating specific locations and categorizing RFOs as non-critical or critical.
\end{itemize}

A detailed description of the formatting, hosting, and licensing details of Hopkins RFOs Bench is in Appendix \ref{dataset_doc}.

\section{Benchmark}
\label{benchmark}

In addition to the Hopkins RFOs Bench dataset, we developed a benchmark for evaluating RFOs detection models on
our cohort. %The code for this benchmark is included in the Github \url{https://github.com/YuliWanghust/RFO_Bench} under an open-source license.

% \subsection{Task Definition}

% We define two primary tasks for evaluation, each associated with specific prediction outputs and ground-truth labels:

% \textbf{Image-Level Classification:}  
% Given an input chest X-ray image $I$, the objective is to predict a binary label $y \in \{0, 1\}$, where $y=1$ indicates the presence of a radiopaque foreign object (RFOs) and $y=0$ denotes its absence. The model learns a mapping function $f_{\text{cls}}: I \rightarrow [0,1]$ that outputs a probability score, which can be thresholded to determine the binary classification.

% \textbf{Object-Level Localization:}  
% For an input image $I$, the goal is to predict a set of $N$ localized objects $\{(b_i, c_i)\}_{i=1}^{N}$, where $b_i$ represents the bounding box or polygon coordinates of the $i$-th detected object, and $c_i \in \{\text{non-interest}, \text{suspicious}, \text{critical}\}$ is the corresponding category label. Ground-truth annotations are provided in the form of bounding boxes or polygons, and evaluation metrics assess both detection accuracy and classification into the appropriate RFOs severity category.

\subsection{Task Definition}

We define two primary tasks for evaluation, each associated with specific prediction outputs and ground-truth annotations.

\textbf{Image-Level Classification:}  
Given a chest X-ray image $I$, the goal is to predict a binary label $y \in \{0, 1\}$, where $y = 1$ indicates the presence of RFOs, and $y = 0$ indicates its absence.  
The model learns a mapping function:
\[
f_{\text{cls}}: I \rightarrow [0,1]
\]
Which outputs a probability score. A threshold (e.g., 0.5 in this paper) is applied to this probability to determine the final binary prediction.

\textbf{Object-Level Localization:}  
Given an image $I$, the task is to detect and localize $N$ RFOs, each represented by a tuple $(b_i, c_i)$ for $i = 1, \dots, N$.  
- $b_i \in \mathbb{R}^4$ denotes the bounding box coordinates $(x_{\text{min}}, y_{\text{min}}, x_{\text{max}}, y_{\text{max}})$.  
- $c_i \in \{\text{non-critical}, \text{critical}\}$ indicates the predicted category of the detected object. The model learns a mapping function:
\[
f_{\text{loc}}: I \rightarrow \{(b_i, c_i)\}_{i=1}^{N}
\]
that outputs a set of localized objects along with their assigned categories.  
%Ground-truth annotations are provided either as bounding boxes or as polygons, and evaluation metrics assess localization accuracy (e.g., Free-response Receiver Operating Characteristic (FROC)).

\subsection{Task Evaluations}

We evaluate models' performance on the two defined tasks — image-level classification and object-level localization — using metrics: Area Under the Curve (AUC) and Free-response Receiver Operating Characteristic (FROC)\cite{egan1961operating,bunch1978free}, respectively.

\textbf{Image-Level Classification:}  
Each algorithm generates a prediction file containing, for each image, a probability score $p \in [0,1]$ indicating the likelihood of RFOs presence.  
We evaluate classification performance using AUC, which is widely adopted in medical imaging, such as CheXpert \cite{irvin2019chexpert} and CAMELYON16 \cite{bejnordi2017diagnostic}, and provides a threshold-independent measure of discriminative ability between RFOs-present and RFOs-absent images. Given the balanced nature of our dataset, AUC serves as an appropriate and robust evaluation metric. In addition, we report accuracy (ACC) and false-negative rate (FNR) as supplementary metrics to further characterize model performance.

\textbf{Object-Level Localization:}  
Each algorithm produces a set of predicted objects, each comprising a bounding box and an associated class label.  
Localization performance is assessed using FROC analysis, which plots sensitivity against the average number of false positives per image.  
Specifically, sensitivity is averaged over a predefined set of false positive rates, offering a holistic view of detection accuracy while balancing true positive and false positive rates.

\section{Baseline Models}

\subsection{Objection Detection Baseline Models} 
We set up several common objection detection modeling approaches for Hopkins RFOs Bench to serve as baselines, including Faster-RCNN \cite{girshick2015fast}, FCOS \cite{tian2020fcos}, RetinaNet \cite{lin2017focal}, and YOLO-v5 \cite{yolov5} models. All details regarding data processing, model training, and evaluation regarding the benchmarks are shown in Appendix \ref{baline_model_details}. As shown in Figure \ref{fig:baselines} (a) illustrates the one-stage or two-stage process of objection detection baseline models encompasses image feature extraction and box classification or regression. Full details, including dataset building, preprocessing, model training and evaluation, and hyperparameter tuning, can be found in Appendix \ref{baline_model_details}.

\begin{figure}[h]
  \centering
    \includegraphics[width=0.76\textwidth]{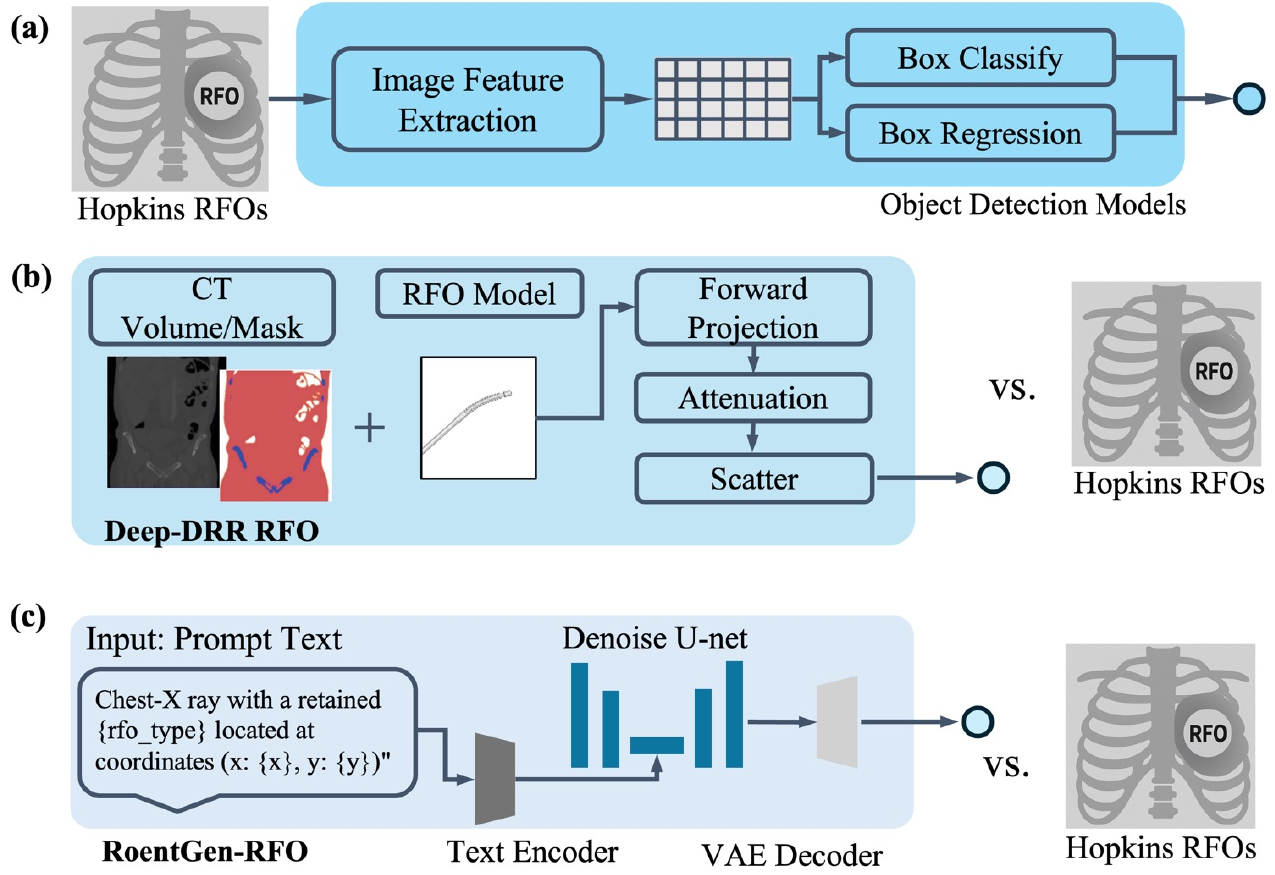}
  \caption{Overview of the baseline frameworks, including: (a) object detection models for RFOs detection, including image feature extraction and box classification or regression, (b) physics-based chest X-ray generation pipeline incorporating RFOs rendering volume, and (c) DDPM-based synthetic chest X-ray generation incorporating prompts specifically for generating critical RFOs.}
  \label{fig:baselines}
  \vspace{-5pt}
\end{figure}

\subsection{Synthetic Chest X-ray Generation}
The detection of RFOs poses a significant challenge due to their rarity in clinical practice, resulting in a data limitation for training deep learning models. To address this, we explore two synthetic chest X-ray generation methods aimed at augmenting RFOs' data availability, particularly for critical RFOs. Specifically, we implement (1) a physics-based generation method using X-ray simulators and (2) deep generative models based on DDPMs. The primary objective is to assess whether synthetic data can mitigate the extreme scarcity of critical RFOs cases and enhance model performance by explicitly incorporating synthetic critical RFOs into datasets.

\textbf{Physics-Based Chest X-ray Generation}
Our proposed physics-based synthetic pipeline, DeepDRR-RFO, is illustrated in Figure~\ref{fig:baselines} (b). The pipeline involves three main steps: (1) CT volume segmentation, (2) RFOs rendering, and (3) physics-based X-ray simulation. First, the CT volumes are segmented into distinct tissue types, including air, soft tissue, and bone. Next, 3D models of RFOs are rendered and spatially integrated into the segmented volumes. Finally, a physics-based digital radiography simulator is used to generate realistic synthetic chest X-rays containing RFOs, accompanied by automatically generated annotations. This approach ensures a physically plausible appearance and anatomical realism of foreign objects within the radiographs. More implementation details are shown in Appendix~\ref{physics_deepdrr}.

\textbf{DDPM-based Chest X-ray Generation.}
Diffusion models, specifically DDPMs, provide an alternative strategy for synthetic image generation. As shown in Figure~\ref{fig:baselines} (c), these models synthesize new images by gradually reversing a noise-injection process, starting from random noise and iteratively denoising to produce high-fidelity chest X-ray images. Unlike physics-based simulations, diffusion models do not require explicit physical modeling of the imaging process and are capable of generating diverse and highly realistic chest X-rays. Using RoentGen-RFO (a diffusion model derived from RoentGen \cite{bluethgen2024vision} to generate datasets containing RFOs, we aim to evaluate their ability to synthesize critical RFOs cases and enrich the training pool with rare but clinically significant examples. More implementation details are shown in Appendix~\ref{ddpm_roentgen}.

\section{Experiments and Results}

\subsection{Baseline Models with Different Training Datasets}

To comprehensively evaluate baseline models' performance and assess the benefit of the Hopkins RFOs Bench, we adopt three different training strategies, with all models evaluated on the Hopkins RFOs Bench testing set. The training strategies include: (1) training on the Object-CXR dataset, (2) training on the Hopkins RFOs Bench, and (3) pretraining on Object-CXR followed by fine-tuning on Hopkins RFOs Bench. Each model is trained for 50 epochs, with an additional 5 epochs for fine-tuning where applicable. Further training details are provided in Appendix~\ref{baline_model_details}. The performance comparison is shown in the following Figure \ref{fig:auc_a}. We notice the prediction performance improved for all baseline models (specifically, Faster-RCNN from 0.62 to 0.80, FCOS from 0.61 to 0.75, Retina from 0.64 to 0.77, and YOLO from 0.61 to 0.78). More results can refer to the Appendix \ref{more_results}.

\begin{figure}[h]
  \centering
    \includegraphics[width=1\textwidth]{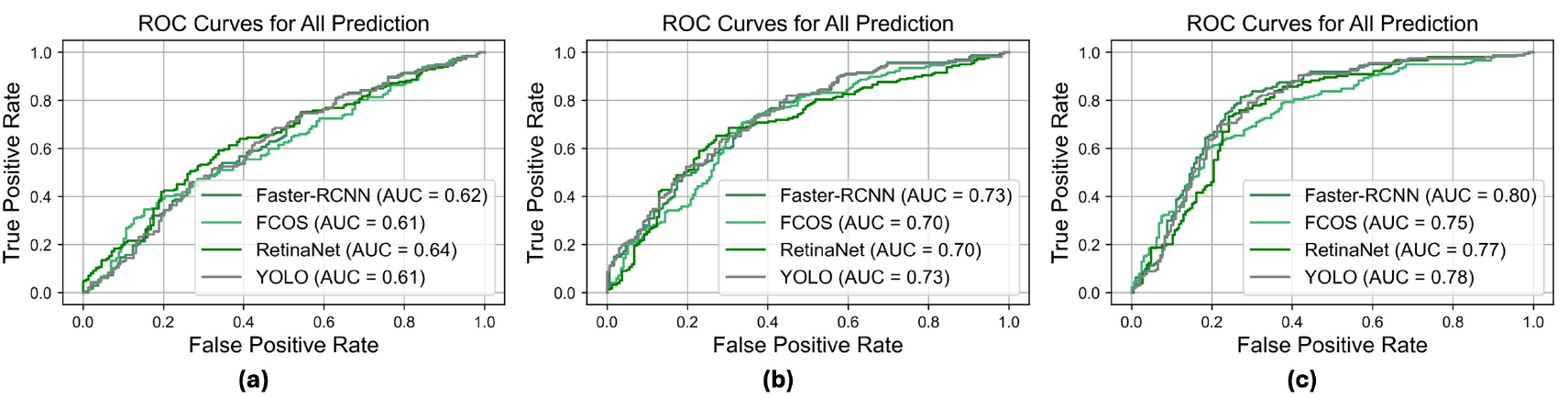}
  \caption{Objection detection performance for (a) training on the Object-CXR dataset, (b) training on the training subset of the Hopkins RFOs Bench, and (c) pretraining on Object-CXR followed by fine-tuning on Hopkins RFOs Bench}
  \label{fig:auc_a}
  \vspace{-5pt}
\end{figure}

\subsection{Baseline Models with Physical-based Synthetic Chest X-ray}

We train the object detection baseline models using four different datasets. The first is the baseline model performance solely using the Hopkins RFOs Bench. In addition, we generate three synthetic datasets using our DeepDRR-RFO pipeline, containing 1,000, 2,000, and 4,000 images, respectively. These synthetic datasets are combined with the Hopkins RFOs Bench dataset to form three distinct training sets, denoted as ‘base+1000’, ‘base+2000’, and ‘base+4000’. For evaluation, all trained models are tested on the held-out testing set of the Hopkins RFOs Bench. Model performance is assessed for two tasks: classification and localization. For classification, we use ACC, FNR, and AUC metrics. Localization performance is evaluated using the FROC metric. 

Full experimental results are summarized in Table~\ref{tab:training_set_results} (a), where it is clear that augmenting the baseline dataset with physics-based synthetic images consistently improves model performance across all metrics, with the best results always observed when adding 2,000 synthetic images. For instance, RetinaNet's performance peaked at 79.5\% ACC, 0.23 FNR, 0.78 AUC, and 63.5 FROC score. Similarly, Faster-RCNN, FCOS, and YOLO exhibited peak performances around the addition of 2,000 physics-based images, after which performance slightly decreased with 4,000 additional synthetic images. This suggests there may be diminishing returns or potential model overfitting when excessively augmenting datasets. More results can refer to the Appendix \ref{more_results}.

\subsection{Baseline Models with DDPM-Based Synthetic Chest X-rays}

We similarly train the baseline models using four datasets incorporating DDPM-based synthetic images. These include the baseline training set of the Hopkins RFOs Bench and four synthetic datasets containing 1,000, 2,000, and 4,000 images. All models are evaluated on the Hopkins RFOs Bench testing set. Results are summarized in Table~\ref{tab:training_set_results} (b). Conversely to the physics-based synthetic chest X-rays, the DDPM-based synthetic image augmentation did not consistently enhance model performance. Across most models and metrics, adding DDPM-based synthetic images generally resulted in decreased or marginally improved performance compared to the baseline. For example, Faster-RCNN demonstrated a decline from baseline (74.0\% ACC, 50.5 FROC) down to 70.0\% ACC and 45.0 FROC score with 4,000 additional DDPM-based images. This highlights critical limitations of DDPM-generated synthetic data in capturing clinically relevant features necessary for robust RFOs detection.

\begin{table*}[t]
\centering
\small
\setlength{\tabcolsep}{4pt}
\renewcommand{\arraystretch}{0.75}
\begin{tabular}{cc}
\toprule
\begin{subtable}[t]{0.48\textwidth}
\centering
\begin{tabular}{llcccc}
\toprule
\textbf{Model} & \textbf{Training Set} & ACC & FNR & AUC & FROC \\
\midrule
Faster & Base & 74.0 & 0.31 & 0.62 & 50.5 \\
-RCNN & Base+1000 & 76.2 & 0.27 & 0.71 & 54.3 \\
& Base+2000 & \textbf{78.5} & \textbf{0.22} & \textbf{0.75} & \textbf{58.7} \\
& Base+4000 & 77.9 & 0.24 & 0.74 & 57.5 \\
\midrule
FCOS & Base & 75.1 & 0.30 & 0.61 & 52.0 \\
& Base+1000 & 76.4 & 0.25 & 0.67 & 55.1 \\
& Base+2000 & \textbf{78.2} & \textbf{0.20} & 0.72 & \textbf{59.8} \\
& Base+4000 & 78.0 & \textbf{0.20} & \textbf{0.74} & 59.5 \\
\midrule
Retina & Base & 74.5 & 0.29 & 0.63 & 53.3 \\
-Net& Base+1000 & 78.0 & 0.26 & 0.77 & 61.7 \\
& Base+2000 & \textbf{79.5} & \textbf{0.23} & \textbf{0.78} & \textbf{63.5} \\
& Base+4000 & 78.7 & 0.24 & 0.76 & 62.1 \\
\midrule
YOLO & Base & 73.8 & 0.33 & 0.62 & 51.2 \\
& Base+1000 & 76.0 & 0.28 & 0.71 & 54.9 \\
& Base+2000 & \textbf{78.7} & \textbf{0.21} & \textbf{0.77} & \textbf{59.7} \\
& Base+4000 & 77.5 & 0.23 & \textbf{0.77} & 57.9 \\
\bottomrule
\end{tabular}
\caption{Physics-based synthetic chest X-rays}
\end{subtable}
&
\begin{subtable}[t]{0.48\textwidth}
\centering
\begin{tabular}{llcccc}
\toprule
\textbf{Model} & \textbf{Training Set} & ACC & FNR & AUC & FROC \\
\midrule
Faster & Base   & \textbf{74.0} & \textbf{0.31} & 0.62 & \textbf{50.5} \\
-RCNN & Base + 1000   & 73.5 & 0.33 & \textbf{0.66} & 48.5 \\
& Base + 2000   & 72.2 & 0.35 & 0.64 & 47.5 \\
& Base + 4000   & 70.0 & 0.38 & 0.61 & 45.0 \\
\midrule
FCOS & Base & \textbf{72.1} & \textbf{0.30} & \textbf{0.61} & \textbf{52.0} \\
& Base + 1000   & 69.0 & 0.37 & 0.54 & 41.0 \\
& Base + 2000   & 70.5 & 0.33 & 0.58 & 43.5 \\
& Base + 4000   & 67.2 & 0.39 & 0.53 & 46.0 \\
\midrule
Retina & Base & \textbf{74.5} & \textbf{0.29} & \textbf{0.63} & \textbf{53.3} \\
-Net& Base + 1000   & 74.0 & 0.31 & 0.66 & 48.0 \\
& Base + 2000   & 71.8 & 0.34 & 0.60 & 44.7 \\
& Base + 4000   & 69.6 & 0.36 & 0.60 & 41.5 \\
\midrule
YOLO & Base & \textbf{73.8} & \textbf{0.33} & \textbf{0.62} & \textbf{51.2} \\
& Base + 1000   & 71.2 & 0.35 & 0.63 & 45.0 \\
& Base + 2000   & 68.3 & 0.38 & 0.58 & 41.4 \\
& Base + 4000   & 69.5 & 0.37 & 0.57 & 42.0 \\
\bottomrule
\end{tabular}
\caption{DDPM-based synthetic chest X-rays}
\end{subtable}
\\
\bottomrule
\end{tabular}
\caption{Experimental results of models trained with different synthetic training sets for (a) Physics-based synthetic chest X-rays and (b) DDPM-based synthetic chest X-rays. For classification, we report the accuracy (ACC), false-negative rate (FNR), and the area under the receiver operating characteristic curve (AUC). Localization performance is evaluated using the Free-response Receiver Operating Characteristic (FROC) metric.}
\label{tab:training_set_results}
\vspace{-8pt}
\end{table*}

\section{Discussion}

Our development of the Hopkins RFOs Bench represents a substantial advancement in medical AI research, specifically addressing the challenging task of detecting critical RFOs on chest X-rays. To our knowledge, Hopkins RFOs Bench is the first publicly available dataset dedicated explicitly to critical RFOs cases. The dataset encompasses carefully annotated chest X-ray images collected from the Johns Hopkins Health System over the past 18 years. %Coupled with our evaluation of leading object detection models—including Faster R-CNN, YOLO, RetinaNet, and FCOS—this resource provides a robust benchmark for the evaluation and improvement of AI-based detection methods for this rare but clinically significant problem. 
Our preliminary findings yield valuable insights into the strengths and limitations of synthetic data augmentation, identifying promising areas for future exploration.

Specifically, we investigated several state-of-the-art object detection models, including Faster R-CNN, YOLO, RetinaNet, and FCOS, each demonstrating relatively fair baseline performance on the Hopkins RFOs Bench. Further analysis indicated that augmenting training datasets with physics-based synthetic images fairly enhanced model performance, particularly when adding around 2,000 synthetic images. All models exhibited improvements in ACC, AUC, and FROC scores with moderate-level synthetic data augmentation. However, performance improvements plateaued or declined when incorporating larger quantities of synthetic images, highlighting potential issues such as overfitting or limitations in synthetic data realism.

While physics-based synthetic images offer valuable training data, they exhibit several limitations. First, synthetic RFOs often appear with lower image resolution and are overly contrasted against the surrounding anatomical background, as illustrated in Appendix Figure~\ref{fig:appen_physics_example}. This visual difference contrasts with the more subtle and variable appearance of RFOs in actual clinical settings. Moreover, the anatomical diversity in these synthetic images is limited by the underlying chest CT volumes used during simulation, restricting the pipeline’s ability to reflect the full range of clinical variability. These limitations constrain the extent to which physics-based synthesis alone can improve model performance in complex, real-world scenarios.

Conversely, DDPM-based synthetic chest X-rays (shown in Appendix Figure \ref{fig:appen_ddpm_example}) offer better visual realism and greater anatomical diversity, unrestricted by the availability of chest CT volumes. Despite this visual quality advantage, current DDPM methods struggle to generate realistic critical RFOs images due to a lack of sufficient real-world RFOs examples necessary for model fine-tuning. Consequently, DDPM-generated images offer limited practical utility for training robust critical RFOs detection models. Nonetheless, our open-access dataset and comprehensive benchmarks establish foundational resources that could enable targeted fine-tuning of DDPM-based models in the future. This advancement has the potential to substantially improve synthetic data generation capabilities and facilitate broader applications in clinical AI research for rare and critical medical conditions.

\textbf{Societal Implications:} Releasing medical imaging datasets, particularly those involving rare and critical findings such as RFOs, involves inherent risks related to patient privacy. To mitigate these concerns, we have strictly followed HIPAA-compliant de-identification procedures and institutional review protocols. Currently, we emphasize that the dataset and accompanying model, while valuable for advancing AI-based research, are not intended for direct clinical use or medical decision-making.

\textbf{Limitation:} Our work has several limitations. First, Hopkins RFOs Bench is derived from a single health system (Johns Hopkins Health System), which may limit generalizability to other institutions or patient populations. Second, while we benchmark existing object detection models on Hopkins RFOs Bench to establish baseline performance, these models are not optimized for critical RFOs detection, underscoring the need for more specialized detection methods in the future. Third, due to the rarity of critical RFOs cases, our dataset remains relatively small, which poses challenges for training high-capacity models. To address this, we developed and evaluated two synthetic data generation methods, DeepDRR-RFO and RoentGen-RFO, yet synthetic images may still lack certain clinical subtleties found in real data. Lastly, annotations in Hopkins RFOs Bench were generated with radiologist oversight, but further manual review and expansion could enhance dataset robustness in future iterations.

\section{Conclusion}

This work makes two primary contributions. First, we introduce Hopkins RFOs Bench, a curated dataset of 144 chest X-ray images containing critical retained foreign objects, collected from the Johns Hopkins Health System over the past 18 years. To our knowledge, this is the largest publicly available dataset focused specifically on critical RFOs. Second, we establish a benchmark by evaluating state-of-the-art object detection models on Hopkins RFOs Bench, highlighting the challenges posed by this rare but clinically significant task and the pressing need for more robust detection methods.

In addition, we assess the utility of synthetic data generation by developing two customized pipelines—DeepDRR-RFO and RoentGen-RFO—for creating chest X-rays with embedded critical RFOs. We train detection models on these synthetic datasets and systematically compare their performance to real-data baselines, providing actionable insights into the strengths and limitations of each approach. By openly sharing Hopkins RFOs Bench and our synthetic generation tools, we aim to catalyze future research on detecting rare findings in radiography and improving model robustness through multimodal and synthetic data augmentation strategies.

% \begin{ack}
% Use unnumbered first level headings for the acknowledgments. All acknowledgments

% Do {\bf not} include this section in the anonymized submission, only in the final paper. You can use the \texttt{ack} environment provided in the style file to automatically hide this section in the anonymized submission.
% \end{ack}

% \section*{References}
\newpage
\makeatletter
\patchcmd{\thebibliography}{\chapter*}{\section*}{}{}
\patchcmd{\thebibliography}{\leftmargin\labelsep}{\leftmargin\labelsep\itemsep=0pt\parskip=0pt plus 1pt\relax}{}{}
\apptocmd{\thebibliography}{\small}{}{}
\makeatother

\bibliographystyle{plain}
\bibliography{mybibliography}

\begin{thebibliography}{10}

\bibitem{andriiashenct}
V~Andriiashen, R~Liere, T~Leeuwen, and KJ~Batenburg.
\newblock Ct-based data generation for foreign object detection on a single x-ray projection. sci. rep. 13 (1), 1881 (2023).

\bibitem{asiyanbola2012modified}
Bolanle Asiyanbola, Chao Cheng-Wu, Jonathan~S Lewin, and Ralph Etienne-Cummings.
\newblock Modified map-seeking circuit: Use of computer-aided detection in locating postoperative retained foreign bodies.
\newblock {\em Journal of Surgical Research}, 175(2):e47--e52, 2012.

\bibitem{bejnordi2017diagnostic}
Babak~Ehteshami Bejnordi, Mitko Veta, Paul~Johannes Van~Diest, Bram Van~Ginneken, Nico Karssemeijer, Geert Litjens, Jeroen~AWM Van Der~Laak, Meyke Hermsen, Quirine~F Manson, Maschenka Balkenhol, et~al.
\newblock Diagnostic assessment of deep learning algorithms for detection of lymph node metastases in women with breast cancer.
\newblock {\em Jama}, 318(22):2199--2210, 2017.

\bibitem{bluethgen2024vision}
Christian Bluethgen, Pierre Chambon, Jean-Benoit Delbrouck, Rogier van~der Sluijs, Ma{\l}gorzata Po{\l}acin, Juan~Manuel Zambrano~Chaves, Tanishq~Mathew Abraham, Shivanshu Purohit, Curtis~P Langlotz, and Akshay~S Chaudhari.
\newblock A vision--language foundation model for the generation of realistic chest x-ray images.
\newblock {\em Nature Biomedical Engineering}, pages 1--13, 2024.

\bibitem{bunch1978free}
PC~Bunch.
\newblock Free response approach to measurement and characterization of radiographic observer performance.
\newblock {\em AJR Am J Roentgenol}, 130(2):382, 1978.

\bibitem{cima2008incidence}
Robert~R Cima, Anantha Kollengode, Janice Garnatz, Amy Storsveen, Cheryl Weisbrod, and Claude Deschamps.
\newblock Incidence and characteristics of potential and actual retained foreign object events in surgical patients.
\newblock {\em Journal of the American College of Surgeons}, 207(1):80--87, 2008.

\bibitem{egan1961operating}
James~P Egan, Gordon~Z Greenberg, and Arthur~I Schulman.
\newblock Operating characteristics, signal detectability, and the method of free response.
\newblock {\em The Journal of the Acoustical Society of America}, 33(8):993--1007, 1961.

\bibitem{gibbs2007preventable}
Verna~C Gibbs, Fergus~D Coakley, and H~David Reines.
\newblock Preventable errors in the operating room: retained foreign bodies after surgery—part i.
\newblock {\em Current problems in surgery}, 44(5):281--337, 2007.

\bibitem{girshick2015fast}
Ross Girshick.
\newblock Fast r-cnn.
\newblock In {\em Proceedings of the IEEE international conference on computer vision}, pages 1440--1448, 2015.

\bibitem{hogeweg2013foreign}
Laurens Hogeweg, Clara~I S{\'a}nchez, Jaime Melendez, Pragnya Maduskar, Alistair Story, Andrew Hayward, and Bram Van~Ginneken.
\newblock Foreign object detection and removal to improve automated analysis of chest radiographs.
\newblock {\em Medical physics}, 40(7):071901, 2013.

\bibitem{hosseini2025synthetic}
Abdullah Hosseini and Ahmed Serag.
\newblock Is synthetic data generation effective in maintaining clinical biomarkers? investigating diffusion models across diverse imaging modalities.
\newblock {\em Frontiers in Artificial Intelligence}, 7:1454441, 2025.

\bibitem{huijben2024denoising}
Evi~MC Huijben, Josien~PW Pluim, and Maureen~AJM van Eijnatten.
\newblock Denoising diffusion probabilistic models for addressing data limitations in chest x-ray classification.
\newblock {\em Informatics in Medicine Unlocked}, 50:101575, 2024.

\bibitem{irvin2019chexpert}
Jeremy Irvin, Pranav Rajpurkar, Michael Ko, Yifan Yu, Silviana Ciurea-Ilcus, Chris Chute, Henrik Marklund, Behzad Haghgoo, Robyn Ball, Katie Shpanskaya, et~al.
\newblock Chexpert: A large chest radiograph dataset with uncertainty labels and expert comparison.
\newblock In {\em Proceedings of the AAAI conference on artificial intelligence}, volume~33, pages 590--597, 2019.

\bibitem{objectcxr}
{JF Healthcare}.
\newblock Object-cxr: Automatic detection of foreign objects on chest x-rays, 2020.
\newblock Accessed: 2025-04-22.

\bibitem{yolov5}
Glenn Jocher.
\newblock Yolov5 by ultralytics, 2020.

\bibitem{johnshopkins_ccda}
{Johns Hopkins University}.
\newblock Center for clinical data analysis (ccda), 2024.
\newblock Accessed: 2025-04-28.

\bibitem{kawakubo2023deep}
Masateru Kawakubo, Hiroto Waki, Takashi Shirasaka, Tsukasa Kojima, Ryoji Mikayama, Hiroshi Hamasaki, Hiroshi Akamine, Toyoyuki Kato, Shingo Baba, Shin Ushiro, et~al.
\newblock A deep learning model based on fusion images of chest radiography and x-ray sponge images supports human visual characteristics of retained surgical items detection.
\newblock {\em International Journal of Computer Assisted Radiology and Surgery}, 18(8):1459--1467, 2023.

\bibitem{khosravi2024synthetically}
Bardia Khosravi, Frank Li, Theo Dapamede, Pouria Rouzrokh, Cooper~U Gamble, Hari~M Trivedi, Cody~C Wyles, Andrew~B Sellergren, Saptarshi Purkayastha, Bradley~J Erickson, et~al.
\newblock Synthetically enhanced: unveiling synthetic data's potential in medical imaging research.
\newblock {\em EBioMedicine}, 104, 2024.

\bibitem{kufel2023chest}
Jakub Kufel, Katarzyna Bargie{\l}-{\L}{\k{a}}czek, Maciej Ko{\'z}lik, {\L}ukasz Czogalik, Piotr Dudek, Miko{\l}aj Magiera, Wiktoria Bartnikowska, Anna Lis, Iga Paszkiewicz, Szymon Kocot, et~al.
\newblock Chest x-ray foreign objects detection using artificial intelligence.
\newblock {\em Journal of Clinical Medicine}, 12(18):5841, 2023.

\bibitem{kumar2017imaging}
GV~Santhosh Kumar, Subhash Ramani, Abhishek Mahajan, Nikshita Jain, Rachel Sequeira, and Meenakshi Thakur.
\newblock Imaging of retained surgical items: A pictorial review including new innovations.
\newblock {\em Indian Journal of Radiology and Imaging}, 27(03):354--361, 2017.

\bibitem{lin2017focal}
Tsung-Yi Lin, Priya Goyal, Ross Girshick, Kaiming He, and Piotr Doll{\'a}r.
\newblock Focal loss for dense object detection.
\newblock In {\em Proceedings of the IEEE international conference on computer vision}, pages 2980--2988, 2017.

\bibitem{lincourt2007retained}
Amy~E Lincourt, Andrew Harrell, Joseph Cristiano, Cathy Sechrist, Kent Kercher, and B~Todd Heniford.
\newblock Retained foreign bodies after surgery.
\newblock {\em Journal of Surgical Research}, 138(2):170--174, 2007.

\bibitem{mahaulpatha2024ddpm}
Praveen Mahaulpatha, Thulana Abeywardane, and Tomson George.
\newblock Ddpm based x-ray image synthesizer.
\newblock {\em arXiv preprint arXiv:2401.01539}, 2024.

\bibitem{mariyaselvam2022central}
Maryanne~ZA Mariyaselvam, Vikesh Patel, Holly~E Young, Mark~C Blunt, and Peter~J Young.
\newblock Central venous catheter guidewire retention: lessons from england’s never event database.
\newblock {\em Journal of Patient Safety}, 18(2):e387--e392, 2022.

\bibitem{o2003imaging}
Angus~R O'Connor, Fergus~V Coakley, Maxwell~V Meng, and Stephen Eberhardt.
\newblock Imaging of retained surgical sponges in the abdomen and pelvis.
\newblock {\em American journal of roentgenology}, 180(2):481--489, 2003.

\bibitem{santosh2021generic}
KC~Santosh, Shotabdi Roy, and Siva Allu.
\newblock Generic foreign object detection in chest x-rays.
\newblock In {\em International Conference on Recent Trends in Image Processing and Pattern Recognition}, pages 93--104. Springer, 2021.

\bibitem{takahashi2023characteristics}
Kyosuke Takahashi, Takeshi Fukatsu, Sayaka Oki, Yusuke Iizuka, Yuji Otsuka, Masamitsu Sanui, and Alan~Kawarai Lefor.
\newblock Characteristics of retained foreign bodies and near-miss events in the operating room: a ten-year experience at one institution.
\newblock {\em Journal of Anesthesia}, 37(1):49--55, 2023.

\bibitem{tian2020fcos}
Zhi Tian, Chunhua Shen, Hao Chen, and Tong He.
\newblock Fcos: A simple and strong anchor-free object detector.
\newblock {\em IEEE transactions on pattern analysis and machine intelligence}, 44(4):1922--1933, 2020.

\bibitem{tochilkin2024triposr}
Dmitry Tochilkin, David Pankratz, Zexiang Liu, Zixuan Huang, Adam Letts, Yangguang Li, Ding Liang, Christian Laforte, Varun Jampani, and Yan-Pei Cao.
\newblock Triposr: Fast 3d object reconstruction from a single image.
\newblock {\em arXiv preprint arXiv:2403.02151}, 2024.

\bibitem{toruner2024artificial}
Merih~Deniz Toruner, Yuli Wang, Zhicheng Jiao, and Harrison Bai.
\newblock Artificial intelligence in radiology: where are we going?
\newblock {\em EBioMedicine}, 109, 2024.

\bibitem{unberath2019enabling}
Mathias Unberath, Jan-Nico Zaech, Cong Gao, Bastian Bier, Florian Goldmann, Sing~Chun Lee, Javad Fotouhi, Russell Taylor, Mehran Armand, and Nassir Navab.
\newblock Enabling machine learning in x-ray-based procedures via realistic simulation of image formation.
\newblock {\em International journal of computer assisted radiology and surgery}, 14:1517--1528, 2019.

\bibitem{unberath2018deepdrr}
Mathias Unberath, Jan-Nico Zaech, Sing~Chun Lee, Bastian Bier, Javad Fotouhi, Mehran Armand, and Nassir Navab.
\newblock Deepdrr--a catalyst for machine learning in fluoroscopy-guided procedures.
\newblock In {\em Medical Image Computing and Computer Assisted Intervention--MICCAI 2018: 21st International Conference, Granada, Spain, September 16-20, 2018, Proceedings, Part IV 11}, pages 98--106. Springer, 2018.

\bibitem{vannucci2013retained}
Andrea Vannucci, Alicia Jeffcoat, Catherine Ifune, Christian Salinas, James~R Duncan, and Michael Wall.
\newblock Retained guidewires after intraoperative placement of central venous catheters.
\newblock {\em Anesthesia \& Analgesia}, 117(1):102--108, 2013.

\bibitem{wang2017chestx}
Xiaosong Wang, Yifan Peng, Le~Lu, Zhiyong Lu, Mohammadhadi Bagheri, and Ronald~M Summers.
\newblock Chestx-ray8: Hospital-scale chest x-ray database and benchmarks on weakly-supervised classification and localization of common thorax diseases.
\newblock In {\em Proceedings of the IEEE conference on computer vision and pattern recognition}, pages 2097--2106, 2017.

\bibitem{wang2024enhancing}
Yuli Wang, Yuwei Dai, Craig Jones, Haris Sair, Jinglai Shen, Nicolas Loizou, Wen-Chi Hsu, Maliha Imami, Zhicheng Jiao, Paul Zhang, et~al.
\newblock Enhancing vision-language models for medical imaging: bridging the 3d gap with innovative slice selection.
\newblock {\em Advances in Neural Information Processing Systems}, 37:99947--99964, 2024.

\bibitem{wang2024car}
Yuli Wang, Wen-Chi Hsu, Victoria Shi, Gigin Lin, Cheng~Ting Lin, Xue Feng, and Harrison Bai.
\newblock Car-dcros: A dataset and benchmark for enhancing cardiovascular artery segmentation through disconnected components repair and open curve snake.
\newblock In {\em International Conference on Medical Image Computing and Computer-Assisted Intervention}, pages 179--189. Springer, 2024.

\bibitem{wasserthal2023totalsegmentator}
Jakob Wasserthal, Hanns-Christian Breit, Manfred~T Meyer, Maurice Pradella, Daniel Hinck, Alexander~W Sauter, Tobias Heye, Daniel~T Boll, Joshy Cyriac, Shan Yang, et~al.
\newblock Totalsegmentator: robust segmentation of 104 anatomic structures in ct images.
\newblock {\em Radiology: Artificial Intelligence}, 5(5), 2023.

\bibitem{webb1988management}
William~A Webb.
\newblock Management of foreign bodies of the upper gastrointestinal tract.
\newblock {\em Gastroenterology}, 94(1):204--216, 1988.

\bibitem{yun2021retained}
Gabin Yun, Ella~A Kazerooni, Elizabeth~M Lee, Palmi~N Shah, Michael Deeb, and Prachi~P Agarwal.
\newblock Retained surgical items at chest imaging.
\newblock {\em Radiographics}, 41(2):E10--E11, 2021.

\bibitem{zejnullahu2017retained}
Valon~A Zejnullahu, Besnik~X Bicaj, Vjosa~A Zejnullahu, and Astrit~R Hamza.
\newblock Retained surgical foreign bodies after surgery.
\newblock {\em Open access Macedonian journal of medical sciences}, 5(1):97, 2017.

\bibitem{zhou2020automatically}
Peng Zhou, Zheng Liu, Hemmings Wu, Yuli Wang, Yong Lei, and Shiva Abbaszadeh.
\newblock Automatically detecting bregma and lambda points in rodent skull anatomy images.
\newblock {\em PloS one}, 15(12):e0244378, 2020.

\end{thebibliography}

\newpage
\appendix

\section{IRB Approval and Data De-identification}\label{IRB}
The development of Hopkins RFO Bench and its associated benchmarks are approved by the Institutional
Review Board IRB of the Johns Hopkins University (JHU). All the data were obtained through the
study "Utility of Artificial Intelligence to Detect Retained Foreign Objects on Radiograph" (IRB Number: IRB00383214). %All patients from Hopkins who are included are going to sign a privacy notice and consent, which informs them that their records may be used for research purposes, given approval by the IRB. %The full dataset will be fully open access \textbf{without restricted access} once the consents are collected, and interested parties can directly use the dataset \textbf{without contacting the Principal Investigator (PI) to obtain the DUA required for full access}. 

Chest radiographs were sourced from the hospital picture archiving and communication system (PACS) in Digital Imaging and Communications in Medicine (DICOM) format. DICOM is a common format that facilitates interoperability between medical imaging devices. The DICOM format contains metadata associated with one or more images, and the DICOM standard stipulates strict rules around the structure of this information. The acquired DICOM images contained PHI ,which required removal for conformance with HIPAA. All Hopkins RFOs Bench are manually reviewed by Hopkins doctors to confirm any protected health information (PHI) is removed from both the DICOM meta-data and the pixel values before public release.

\section{Cohort Definition}

The cohort definition protocol for the Hopkins RFO dataset is illustrated in Figure~\ref{fig:cohort}. With IRB approval, we retrospectively reviewed over 50,000 chest X-ray studies collected from the Johns Hopkins Health System between 2007 and 2024. In the first step, we filtered these cases using keyword-based search in radiology reports, identifying 6,371 cases containing potentially suspicious retained foreign objects (RFOs), flagged by phrases such as “foreign body” or “radio-opaque foreign body,” while excluding negated mentions, such as such as “no foreign body” or “no radio-opaque foreign body”. %All images are included in the Github link \url{https://anonymous.4open.science/r/RFO_Bench-8742/README.md}

Next, board-certified radiologists conducted a detailed inspection of these 6,371 cases, reviewing chest X-rays alongside positive control samples, corresponding radiology reports, and surgical or operational notes. This process identified 144 cases containing clinically significant, critical RFOs. These cases then underwent a formal annotation phase, in which radiologists labeled the images using a standardized annotation protocol, followed by a second independent read to ensure quality and consistency. The result is a curated and expert-annotated dataset of 144 critical RFO cases, referred to as the Hopkins RFO Bench.

\subsection{Dataset Documentation}\label{dataset_doc}

\subsection*{Hosting, Access, License, and Long-Term Preservation}
The Hopkins RFO dataset, including annotated chest X-rays and associated metadata, is hosted by Johns Hopkins University. No Data Use Agreement (DUA) is required for access. The dataset will be made fully and publicly available upon acceptance of the corresponding manuscript. Our lab will ensure long-term preservation and access through institutional hosting infrastructure.

As the authors and curators of the dataset and manuscript, we affirm full responsibility for its contents. All data were collected in compliance with institutional IRB approval and relevant ethical standards. The dataset has undergone manual review to ensure the removal of any patient-identifying information prior to release. We confirm that the dataset does not infringe on any copyright, proprietary, or personal rights, and that all necessary permissions have been secured.

% \subsection*{License Terms of Use}
% The Hopkins RFO dataset will be released under a non-commercial, research-use license to support open research and development. Users may freely access and use the data for academic and research purposes, provided proper citation of the dataset and accompanying publication. Complete licensing terms and citation guidelines will be provided on the dataset website upon release.

\section{Dataset Annotation and Illustration}

\subsection{Dataset Annotation}\label{anno_prot}

\begin{figure}[h]
  \centering
  \includegraphics[width=1\textwidth]{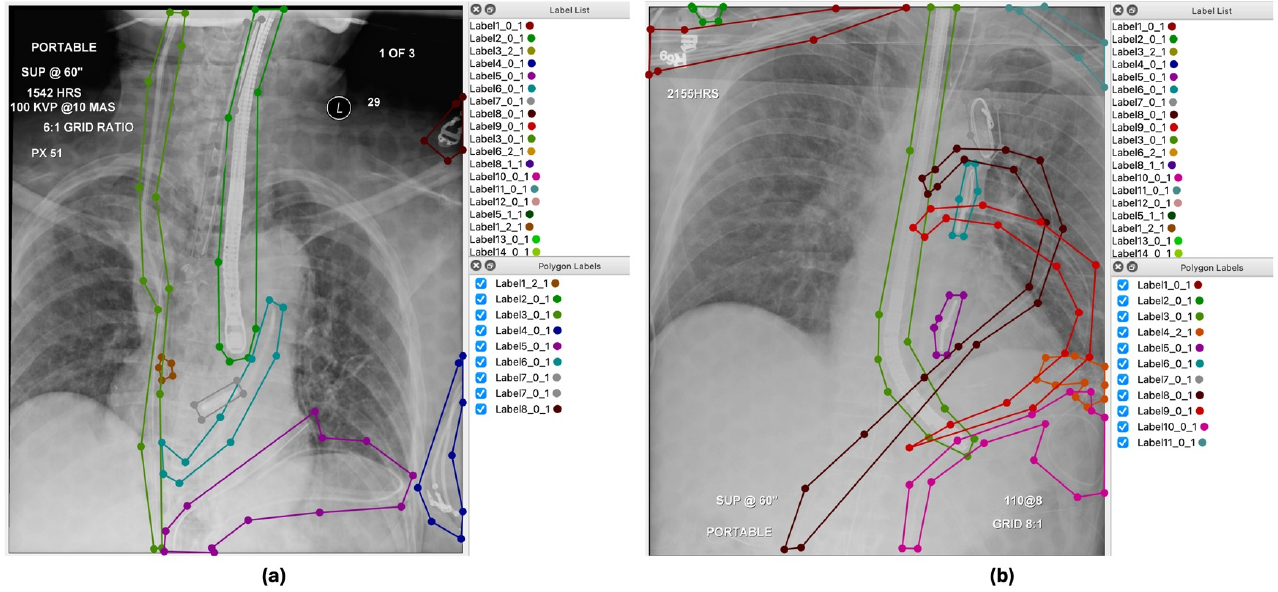}
  \caption{Two examples of radiologist annotations for RFOs using the defined annotation protocol shown below. Each image illustrates the location and type of RFOs. These examples also highlight differences in annotation between critical and non-critical cases.}
  \label{fig:annotation_examples}
\end{figure}

Image-level annotation is provided by organizing images into separate folders: one for RFO-positive cases and another for no-RFO cases.

Object-level annotation for each image is provided, which indicates the accurate location of each foreign object using a closed shape. Annotations are provided in a JSON file, and a JSON example is shown below.

\begin{verbatim}
image_path,annotation
/path/#####.jpg, <ANNO-Number-ID>_<RFO-type>_<ANNO-TYPE-IDX> x1 y1 x2 y2;
<ANNO-Number-ID>_<RFO-type>_<ANNO-TYPE-IDX> x1 y1 x2 y2 ... xn yn;...
/path/#####.jpg,
/path/#####.jpg, <ANNO-Number-ID>_<RFO-type>_<ANNO-TYPE-IDX> x1 y1 x2 y2;
/path/#####.jpg, <ANNO-Number-ID>_<RFO-type>_<ANNO-TYPE-IDX> x1 y1 x2 y2 
... xn yn;...
......

\end{verbatim}
\texttt{ANNO-Number-ID} begins at 1 and increments sequentially to uniquely index each annotated instance, covering all types of RFOs across the dataset.

\texttt{RFO-type} specifies the classification of each RFO instance and includes two categories: critical (encoded as \texttt{1}) and non-critical (encoded as \texttt{0}).

Two types of shapes are used, namely rectangles and polygons. We use \texttt{0} and \texttt{1} as \texttt{ANNO-TYPE-IDX} respectively.

\begin{itemize}
    \item For rectangle annotations, we provide the bounding box (upper left and lower right) coordinates in the format \texttt{x1 y1 x2 y2}, where \texttt{x1 < x2} and \texttt{y1 < y2}.
    \item For polygon annotations, we provide a sequence of coordinates in the format \texttt{x1 y1 x2 y2 ... xn yn}.
\end{itemize}

\textbf{Note:} Our annotations use a Cartesian pixel coordinate system, with the origin (0,0) in the upper left corner. The x coordinate extends from left to right; the y coordinate extends downward.

Two examples of annotated images, along with corresponding explanations, are shown in Figure~\ref{fig:annotation_examples}. These demonstrate how radiologists labeled RFOs across different clinical scenarios.

\subsection{Examples of Hopkins RFOs Bench}

\begin{figure}[h]
  \centering
    \includegraphics[width=1\textwidth]{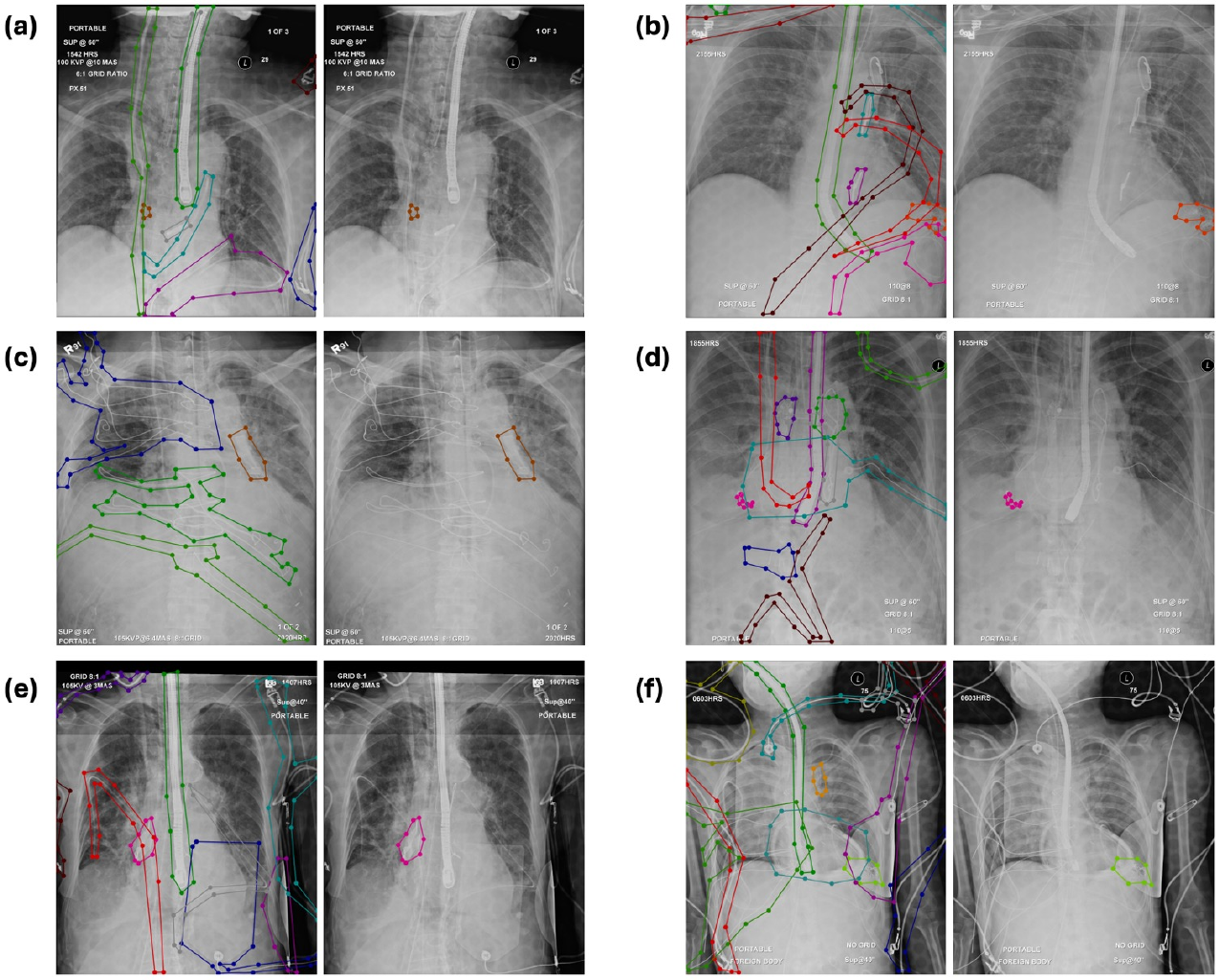}
  \caption{Annotation examples of six unidentified RFO cases, labeled (a) through (f), from six different patients. In each example, the left sub-figure displays all detected RFOs—including non-critical and critical—while the right sub-figure highlights only the critical RFOs.}
  \label{fig:appen_hopkins_example}
\end{figure}

Representative examples from the Hopkins RFOs Bench are illustrated in Figure~\ref{fig:appen_hopkins_example}, showing the diversity and complexity of RFOs cases in real-world clinical settings. The figure presents six anonymized cases, labeled (a) through (f), each from a distinct patient.

These examples highlight the visual heterogeneity of RFO types in chest X-rays, including wires, sutures, needles, and surgical sponges. They also demonstrate the varying anatomical locations and imaging appearances of RFOs, posing challenges for automated detection. All annotations were performed by expert radiologists following a standardized protocol and validated through a second read to ensure high labeling accuracy and clinical relevance.

\section{Objection detection Baseline}\label{baline_model_details}

%The code for baseline models is included in the Github link \url{https://anonymous.4open.science/r/RFO_Bench-8742/README.md}. 
Here are some descriptions for each step we delivered in the code, including data preprocessing, model training, and evaluation:

\textbf{Preprocessing:} To construct and preprocess the dataset for RFOs detection, we implement a custom \texttt{ForeignObjectDataset} class that supports both training and validation modes. The dataset is initialized with a folder of images and a dictionary (\texttt{labels\_dict}) that maps image filenames to their corresponding bounding box annotations. Only images with valid annotations are included. For training samples (\texttt{datatype=‘train'}), annotations are parsed from a semicolon-separated string format, where each entry specifies a bounding box or polygon marking an RFO. The coordinates are normalized and converted into bounding box format $[x_{\min}, y_{\min}, x_{\max}, y_{\max}]$, and scaled to a fixed image resolution of 600~$\times$~600 pixels. Each image is then associated with a label tensor (assigning class label 1 for all RFOs).

For validation data (\texttt{datatype=‘dev'}), binary labels are assigned based on the presence or absence of any annotation. All images undergo a preprocessing pipeline using \texttt{torchvision.transforms}, which includes resizing to 600~$\times$~600 and normalization using ImageNet mean and standard deviation. The processed data is used for training and validation, using batch sizes of 8 and 1, respectively. A custom \texttt{collate\_fn} is used to support variable-length annotations. This setup enables flexible and consistent training of object detection models using both bounding box–level supervision and image-level classification.

\textbf{Model Training:} We trained four object detection models, each adapted for binary classification by modifying the classification head—such as replacing the default predictor with \texttt{FastRCNNPredictor} in the Faster R-CNN architecture—to distinguish between images with and without RFOs. Models were trained using stochastic gradient descent (SGD) with a learning rate of 0.005, momentum of 0.9, and a weight decay of 0.0005. A step-based learning rate scheduler was applied, reducing the learning rate by a factor of 0.1 every 5 epochs.

During training, we evaluated model performance at each epoch using image-level classification metrics. Specifically, we thresholded the maximum predicted object confidence score at 0.5 to generate binary predictions. If no objects were detected in a validation image, a prediction score of 0 was assigned; otherwise, the highest confidence score among detected objects was used. Validation accuracy and AUC were computed, and the model checkpoint with the highest AUC was retained. This training strategy enables robust learning of both object localization and image-level classification, even in the context of sparse and visually diverse RFO presentations.

\section{Image Cohort from DeepDRR-RFO} \label{physics_deepdrr}

\begin{figure}[h]
  \centering
    \includegraphics[width=0.8\textwidth]{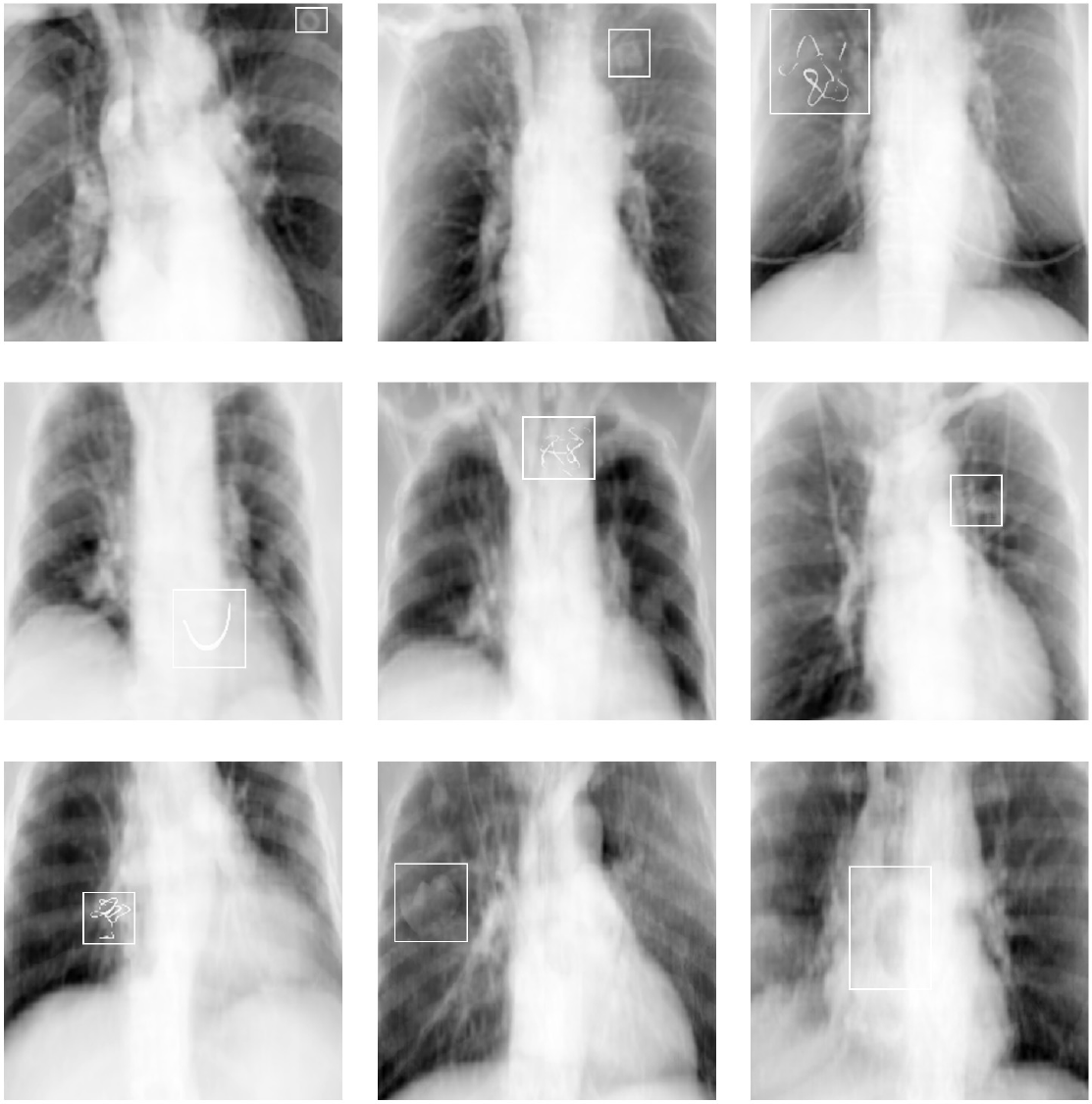}
  \caption{Nine examples of physics-based synthetic chest X-rays with critical RFOs, each generated from a different patient. White bounding boxes indicate the locations of RFOs, including wires, sutures, sponges, and rings.}
  \label{fig:appen_physics_example}
\end{figure}

To address the scarcity of annotated critical RFO cases and overcome the limitations of real-world data collection, we propose to use DeepDRR-RFO, a physics-based pipeline for generating realistic synthetic chest X-ray image with automatically labeled critical RFOs. The pipeline combines high-quality CT segmentation, 3D reconstruction of common surgical/clinical operational items, and X-ray simulation via an optimized DeepDRR framework \cite{unberath2018deepdrr} to produce clinically plausible training data for AI models. As illustrated in Figure~\ref{fig:baselines}, our method integrates four key components:
\begin{itemize}
    \item (1)segmentation of CT volumes into anatomical materials using a deep learning-based tool (e.g., Total Segmentor \cite{wasserthal2023totalsegmentator} in our paper available at \url{https://github.com/wasserth/TotalSegmentator});
    \item (2) construction of 3D RFO models from real surgical items via single-image reconstruction (e.g., Triposor \cite{tochilkin2024triposr} available at \url{https://github.com/VAST-AI-Research/TripoSR});
    \item (3) physics-based X-ray rendering using material-specific attenuation properties (e.g., National Institute of Standards and Technology (NIST) database \url{https://www.nist.gov/data} in this paper);
    \item (4) automated projection of RFO coordinates to generate pixel-level annotations.
\end{itemize}

This pipeline enables the creation of large-scale, diverse, and annotated synthetic datasets with high realism and minimal manual effort, thereby facilitating the development of robust RFO detection algorithms in the absence of abundant real-world critical RFOs data.

To demonstrate the effectiveness of our DeepDRR-RFO pipeline, we present examples of DeepDRR-RFO simulated chest X-rays with critical RFOs in Figure~\ref{fig:appen_physics_example}. Each image corresponds to a different patient CT volume and contains one or more retained foreign objects, such as wires, sutures, sponges, or surgical rings. The white bounding boxes indicate the precise locations of these RFOs, which are automatically annotated during the simulation process. These examples highlight the visual realism of the synthetic radiographs and illustrate the range of object appearances, sizes, and anatomical placements captured by our pipeline.

\begin{figure}[h]
  \centering
    \includegraphics[width=1\textwidth]{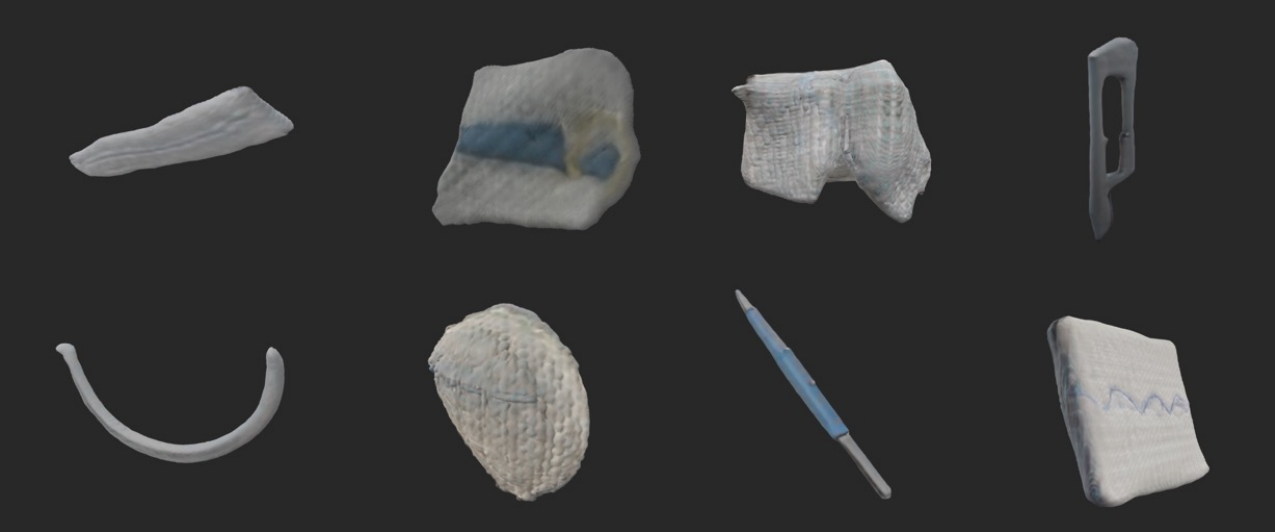}
  \caption{Eight examples of clinical RFO 3D rendering models used for physics-based synthetic chest X-ray generation, representing various object types including wires, sutures, sponges, and rings. The rndring volume of each RFO are open access on our Hopkins RFO Bench.}
  \label{fig:appen_physics_rfo_rendering}
\end{figure}

Figure~\ref{fig:appen_physics_rfo_rendering} further illustrates the underlying 3D RFO models used in our rendering process. This reconstruction uses the TripoSR model from single-view photographs of actual surgical items, embedded into patient CT volumes during simulation. By incorporating a diverse set of clinically relevant RFO types, the pipeline ensures that the synthetic dataset closely reflects real-world variability, which is essential for training generalizable object detection models.

\section{Image Cohort for RoentGen-RFO} \label{ddpm_roentgen}

The \textbf{RoentGen-RFO} model is adapted from \textbf{RoentGen} \cite{bluethgen2024vision}, a state-of-the-art diffusion-based generative framework designed to synthesize realistic chest X-rays. Leveraging DDPMs, RoentGen produces anatomically accurate and diverse synthetic radiographs that closely resemble real clinical images. We obtained access to the model through a signed user agreement with Stanford, and all experiments conducted using RoentGen-RFO are strictly for research purposes and do not provide medical advice.

RoentGen-RFO implements RoentGen in a \textit{zero-shot} manner, without additional training or fine-tuning on RFO-specific datasets, by employing specially designed input prompts designed for RFOs. For example, a structured prompt such as \texttt{"Normal chest X-ray with a retained {rfo-type} located at coordinates {(x: {x}, y: {y})} in the quadrant of the image"} enables the model to generate corresponding synthetic images on demand. Parameters such as the type and location of the foreign object are specified via CSV files, facilitating reproducibility and ease of integration into automated pipelines.

\begin{figure}[h]
  \centering
    \includegraphics[width=1\textwidth]{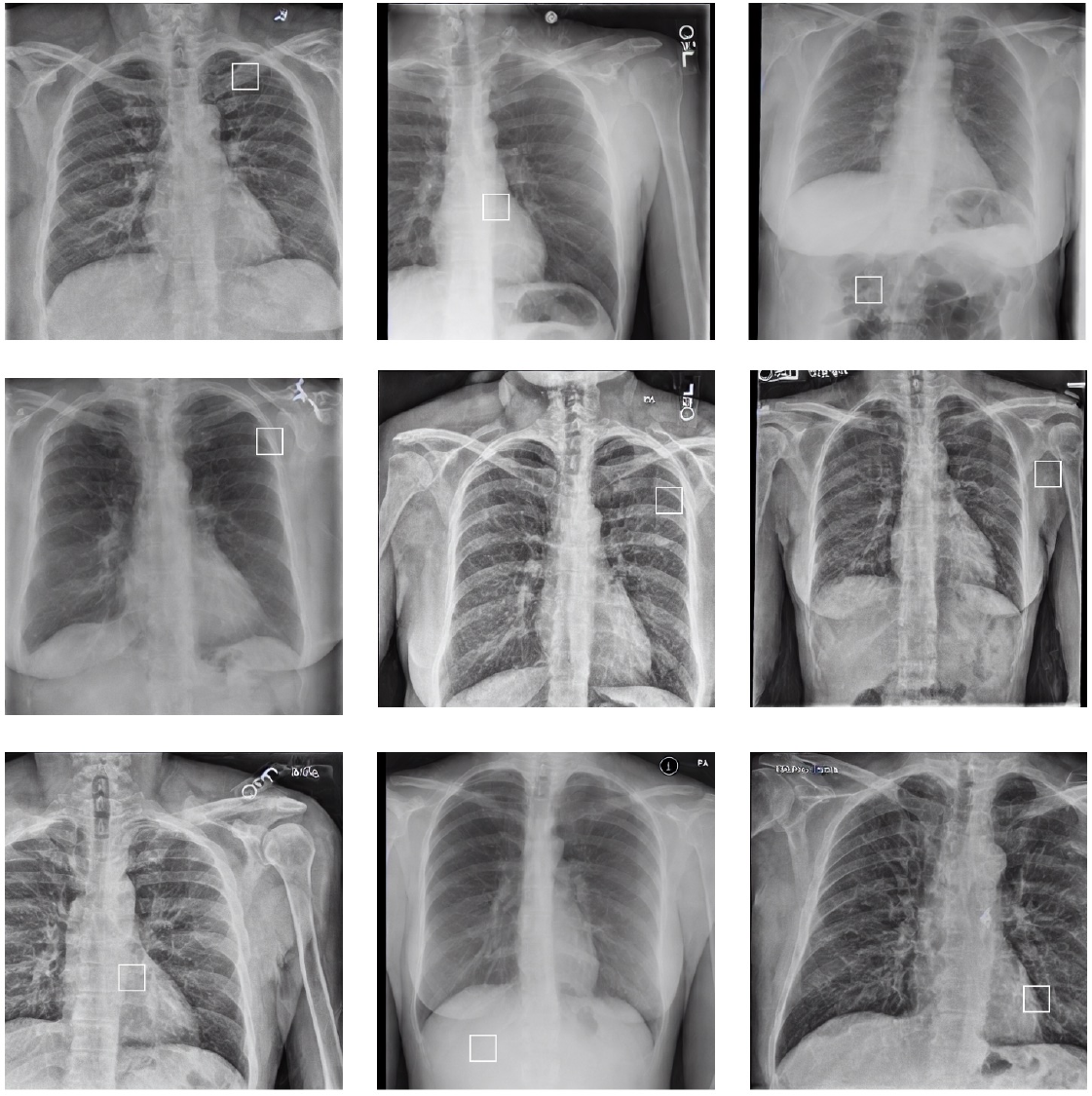}
  \caption{Nine examples of ddpm-based synthetic chest X-rays with critical RFOs, each generated from a different patient. White bounding boxes indicate the locations of RFOs, including wires, sutures, sponges, and rings.}
  \label{fig:appen_ddpm_example}
\end{figure}

Figure~\ref{fig:appen_ddpm_example} illustrates examples of synthetic chest X-rays generated using RoentGen-RFO, showcasing a range of critical RFOs, including wires, sutures, sponges, and rings, and demonstrating the model’s capability to simulate rare yet clinically important scenarios.

\begin{tcolorbox}[colback=white, colframe=black!75!black, width=\textwidth, title=RoentGen-RFO Input Prompt, fonttitle = \bfseries, fontupper=\ttfamily]
\label{prompt::abnormality_extraction}
Normal chest X-ray with a retained \texttt{rfo-type} located at coordinates \texttt{(x: {x}, y: {y})} in the quadrant of the image.
\end{tcolorbox}
\texttt{rfo-type} refers to \texttt{wire, suture, needle, sponge, electrode and so on.}

\texttt{(x: {x}, y: {y})} refers to any location of the generated image. %Both information are saved in a CSV file and open on \url{https://huggingface.co/datasets/Yuliiiiiiiione/Hopkins_RFO_Bench/tree/main/DDPM-based%20sythetic%20images}.

\section{More Results for Baselines, Physics-based synthetic method and DDPM based method}\label{more_results}

To further evaluate baseline object detection models, we report detailed classification and localization results across three different training setups in Table~\ref{tab:different_training_set_results}. These include models trained solely on Object-CXR, on the Hopkins RFOs Bench, and on a combined training scheme that uses Object-CXR for pre-training followed by fine-tuning on the Hopkins RFOs Bench. We benchmark four commonly used detectors—Faster R-CNN, FCOS, RetinaNet, and YOLO—using standard performance metrics to assess both classification (ACC, FNR, AUC) and localization (FROC) capabilities.

\begin{table*}[t]
\centering
\small
\setlength{\tabcolsep}{4pt}
\renewcommand{\arraystretch}{0.75}
\centering
\begin{tabular}{llcccc}
\toprule
\textbf{Training Set} & \textbf{Models} & ACC & FNR & AUC & FROC \\
\midrule
Object-CXR & Faster-RCNN & 69.0 & 0.41 & 0.62 & 38.9 \\
& FCOS & 66.2 & 0.37 & 0.61 & 34.2 \\
& RetinaNet & 68.5 & 0.36 & 0.64 & 37.6 \\
& YOLO & 66.9 & 0.44 & 0.61 & 35.5 \\
\midrule
Hopkins RFOs Bench & Faster-RCNN & 74.3 & 0.29 & 0.73 & 49.8 \\
& FCOS & 71.4 & 0.32 & 0.67 & 45.1 \\
& RetinaNet & 70.2 & 0.29 & 0.70 & 49.0 \\
& YOLO & 75.0 & 0.26 & 0.74 & 50.5 \\
\midrule
Object-CXR + & Faster-RCNN & 74.5 & 0.29 & 0.80 & 53.3 \\
Hopkins RFOs Bench & FCOS & 73.0 & 0.26 & 0.75 & 51.7 \\
& RetinaNet & 73.5 & 0.23 & 0.77 & 53.1 \\
& YOLO & 75.7 & 0.24 & 0.78 & 50.2 \\
\bottomrule
\end{tabular}
\caption{Classification and localization performance of object detection models trained on different synthetic datasets.
We evaluate four models—Faster R-CNN, FCOS, RetinaNet, and YOLO—across three training settings: (a) Object-CXR, (b) Hopkins RFOs Bench, and (c) the combined dataset using Object-CXR for pre-training and using Hopkins RFO Bench for fine-tuning. Metrics reported include accuracy (ACC), false-negative rate (FNR), area under the receiver operating characteristic curve (AUC), and Free-response Receiver Operating Characteristic (FROC) for localization performance.}
\label{tab:different_training_set_results}
\end{table*}

\begin{figure}[h]
  \centering
    \includegraphics[width=0.85\textwidth]{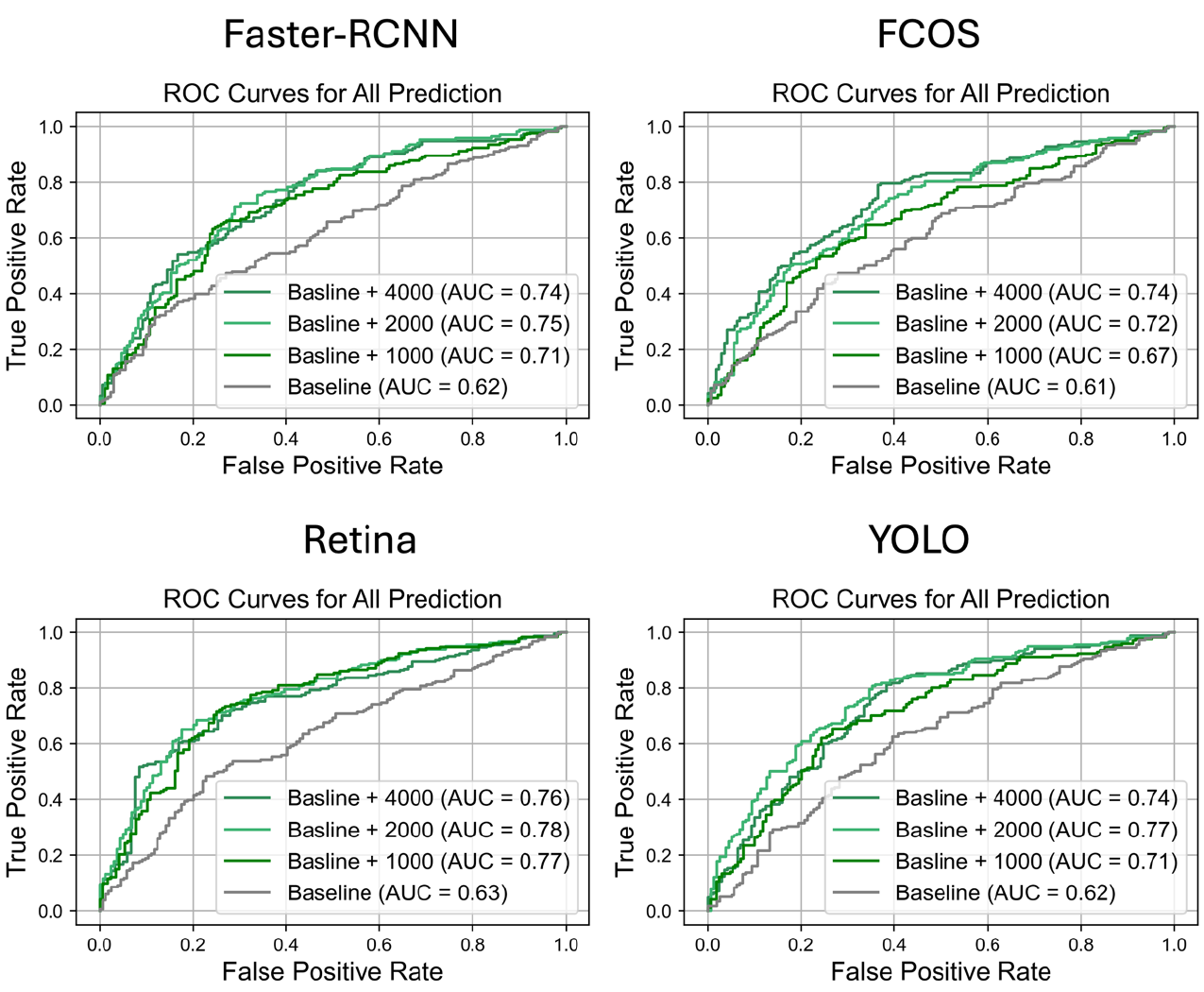}
  \caption{Object detection performance for (a) training on the Hopkins RFO dataset alone, (b) with 1,000 additional Physics-based synthetic chest X-rays, (c) with 2,000 synthetic images, and (d) with 4,000 synthetic images added to the Hopkins RFO dataset.}
  \label{fig:auc_physics}
\end{figure}

\begin{figure}[h]
  \centering
    \includegraphics[width=0.85\textwidth]{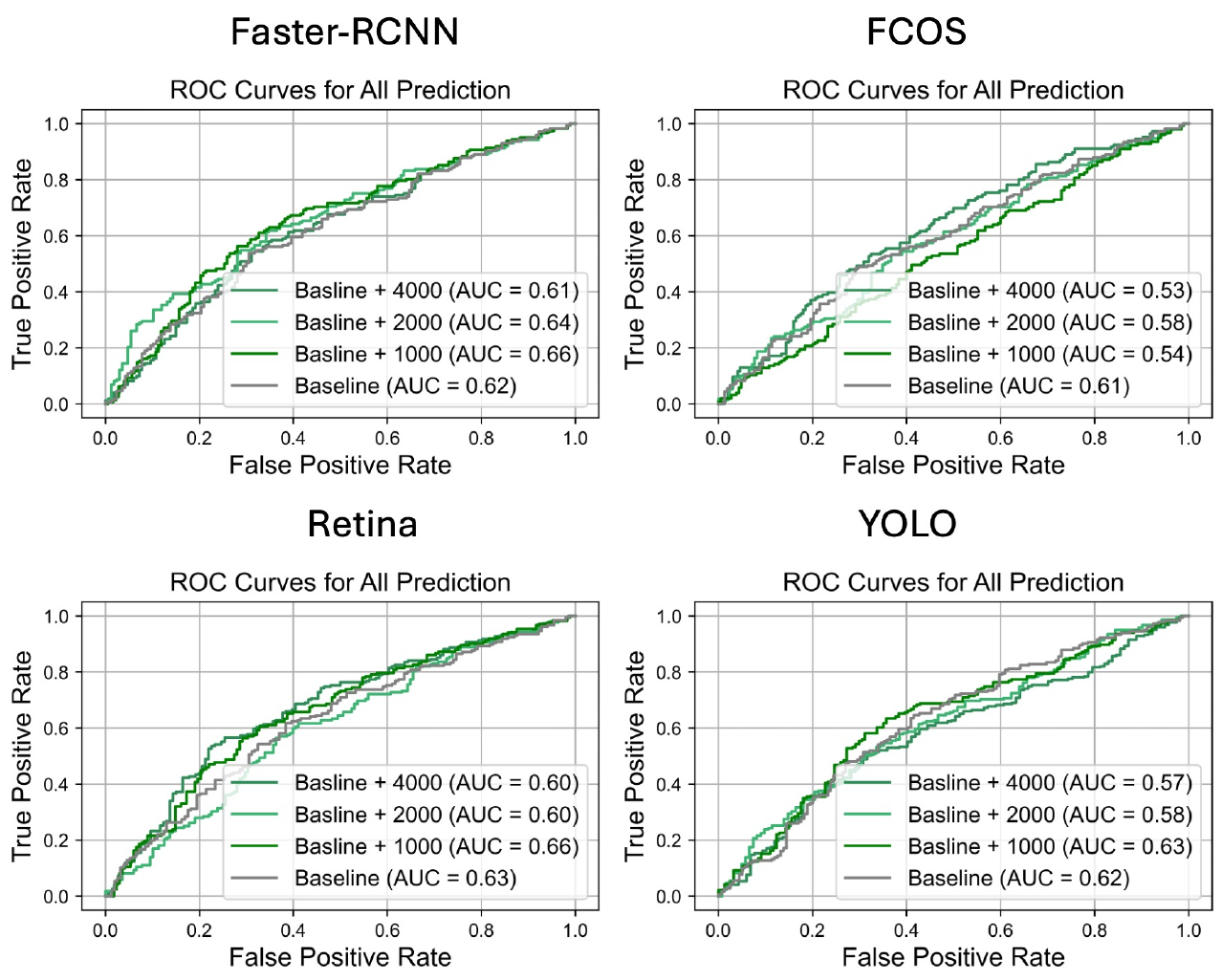}
  \caption{Object detection performance for (a) training on the Hopkins RFO dataset alone, (b) with 1,000 additional DDPM-based synthetic chest X-rays, (c) with 2,000 synthetic images, and (d) with 4,000 synthetic images added to the Hopkins RFO dataset.}
  \label{fig:auc_ddpm}
\end{figure}

Figure~\ref{fig:auc_physics} and Figure~\ref{fig:auc_ddpm} illustrate the impact of augmenting the Hopkins RFO dataset with synthetic chest X-rays on object detection performance. In these experiments, synthetic data was generated using two different approaches: Physics-based synthetic method and deep generative modeling via diffusion (DDPM) based method. Each figure compares model performance when trained on the real dataset alone versus when progressively larger amounts of synthetic data (1,000, 2,000, and 4,000 images) are added.

Figure~\ref{fig:auc_physics} shows the results of incorporating physics-based synthetic images into the training process. As additional synthetic samples are introduced, the model demonstrates incremental performance gains, suggesting that the simulated data successfully enriches the representation of RFO-relevant features. This trend highlights the utility of domain-informed image synthesis techniques in addressing data scarcity while maintaining physical plausibility.

Conversely to the physics-based synthetic chest X-rays, the DDPM-based synthetic image augmentation did not consistently enhance model performance, as shown in Figure~\ref{fig:auc_ddpm}. Across most models and metrics, adding DDPM-based synthetic images generally resulted in decreased or marginally improved performance compared to the baseline.

\end{document}